\documentclass[12pt]{elsarticle}

\makeatletter
\def\ps@pprintTitle{%
 \let\@oddhead\@empty
 \let\@evenhead\@empty
 \def\@oddfoot{}%
 \let\@evenfoot\@oddfoot}
\makeatother

\usepackage{amssymb}

\usepackage{lineno}

\usepackage{mwe}
\usepackage{subfig}
\usepackage{float}
\usepackage{indentfirst}
\usepackage[export]{adjustbox}
\usepackage{tabu}
\usepackage{graphicx}
\usepackage{hyperref}
\usepackage{fancyhdr}
\usepackage{textpos}

\journal{Journal Name}

\begin{document}

\begin{frontmatter}



\title{Investigation of the thermal and hydraulic properties of a novel heat exchanger for smart heat supply systems}



\author[add1,add2]{Dmitry Loginov}
\author[add1]{Alexander Ustinov}

\address[add1]{Skolkovo Institute of Science and Technology, Moscow, Russia}
\address[add2]{Moscow Institute of Science and Technology, Dolgoprudny, Russia}


\begin{abstract}
In this study, analysis of shell and tube heat exchanger (HE) is performed. Theory part on heat transfer, calculation of heat exchanger and general thermal and hydrological properties are described. Several models are developed and computed in different cases: plain tubes and twisted tubes; common and separated inlet and outlet; heat exchanger with different baffles and geometry inside a shell; changed gas matter, tube's material, tube's thickness; modification of inlet and outlet. Twisted tubes shows better heat transfer efficiency than plain tubes. Baffles increase intensity of water mixing but dead zones can be formed in some cases. Length of air inlet and outlet influence on temperature distribution that is important for location of measuring systems. Experimental setup for a gas-liquid case is being created to verify the modelling and test new heat exchanger. 
\end{abstract}


\end{frontmatter}


\section{Introduction}
\label{S:1} 
Nowadays heat exchanging systems are used in various areas \cite{App1}, \cite{App2}: utility services, cars, chemical industry, factories, aircrafts, space systems, etc. New materials, technologies and designs allow to increase the efficiency of a device, reduce heat and energy losses and undertake optimization study of heat exchangers. The increase in heat transfer will lead to reduction system’s cost and size, that is very important in different fields, like car industry and Smart District Heating, as there are increased requirements for heat exchanger on the efficiency in heat pump system \cite{SDH2}, \cite{SDH1}. \par 

\begin{textblock*}{200mm}(0\textwidth,1.8 cm)
\noindent \underline{Email addresses}: \\ \textbf{dmitry.loginov@skolkovotech.ru (Dmitry Loginov)}, \\ alexander.ustinov@gmail.com (Alexander Ustinov).
\end{textblock*}

As it is high interest in heat exchanger from an industry side and growth research of Smart Grids and Smart Cities topics, where efficiency of heat exchanger in heat pump is extremely important it was decided to get into the theme. To evaluate numerically efficiency of a shell and tube heat exchanger and its thermal and hydrological properties CFD model should be created. After that the system is needed to modify to get higher efficiency and analyze HE in different cases, like twisted tubes, baffles location, geometry changes, parallel flow and counter flow. After that setup for testing heat exchanger should be created to compare computational results with theory and experimental data.
The final goal of the project is to develop an experimental setup for testing and certification different types of heat exchangers both industrial and new designed prototypes. 

\section{Calculation}

The purpose of the heat calculation of the heat exchanger is to determine the heat fluxes in the apparatus, its heat load, the true values of the heat transfer coefficients, the wall temperatures and the heat exchange surface.

The heat balance equation has the form \cite{4712}: \\

\begin{equation} 
Q' = G_1 \ \Delta i_1 = G_2 \ \Delta i_2
\end{equation}
where: \\
$Q'$ - thermal power of a heat exchanger, $W$; \\
$G_1, \ G_2$ - coolant mass flow, $kg/s$; \\
$\Delta i_1, \ \Delta i_1$ - enthalpy change of heat carries, $J/kg$ \\

The thermal load of the apparatus is determined from the equation of the thermal balance by the equation:\\

\begin{equation} 
Q'=\frac{Q}{\tau} = G \ c \ (T_{in} - T_{out})
\end{equation}
where:\\
$Q'$  - thermal power, $W$; \\
$Q$ - amount of heat, $J$; \\
$\tau$ - time, $s$; \\
$G$ - mass flow rate of the second heat carrier fluid, $kg / s$; \\
$c$ - Specific heat of the liquid, $J / (kg \cdot K)$; \\
$T_{out}$ - final temperature of the fluid, $K$; \\
$T_{in}$ -  initial temperature of the fluid, $K$. \\

\bigskip

The amount of heat transferred from the hot coolant to the cold one through the wall separating them is determined by the \textbf{heat transfer equation}:\\

\begin{equation} 
Q=k \ F \ \Delta T_{av} \ \tau
\end{equation}
where: \\
$Q$ -  the amount of transferred heat, $J$; \\
$k$ - coefficient of heat transfer, ${W}/{(m^2 \cdot K)}$;\\
$F$ - heat transfer surface, $m^2$;\\
$\Delta T_{av}$ - average temperature difference between hot and cold coolant, $K$;\\
$\tau$ - is the time, $s$. \\

The heat flow in the apparatus or the thermal power of the apparatus is determined by the equation:\\

\begin{equation} 
Q'=\frac{Q}{\tau} = k \ F \ \Delta T_{av}
\end{equation}

\begin{equation} 
Q = k \ F \ \Delta T_{av}
\end{equation}

And the specific heat flux:

\begin{equation} 
q=\frac{Q}{F \tau} = k \ \Delta T_{av}
\end{equation}

The \textbf{heat transfer coefficient} for a flat heat transfer surface and for a cylindrical heat transfer surface at $\frac{\delta_{w}}{d_{in}} < 0,3-0,4 $ (where: $\delta_{w}$ - is the wall thickness, $m$, $d_{in}$ - internal diameter of the cylindrical surface, $m$) is calculated by the formula:\\
 
\begin{equation} 
k=\frac{1}{\frac{1}{\alpha_1} + \frac{\delta_w}{\lambda_w} + \sum r + \frac{1}{\alpha_2} }
\end{equation}
where: \\
$\alpha_1$ - the heat transfer coefficient for the hot coolant, $W / (m^2 \cdot K)$;\\
$\lambda_w$ - coefficient of thermal conductivity of the wall material, $W / (m \cdot  K)$;\\
$\sum r$ - the sum of the thermal resistances of scale and wall contaminants, $m^2 \cdot K / W$;\\
$\alpha_2$ - the heat transfer coefficient for the cold coolant, $W / (m^2 \cdot K)$.\\

From the basic heat transfer equation, the heat exchange surface is determined by:\\

\begin{equation} 
F = \frac{Q}{k \ \Delta T_{av} \ \tau} = \frac{Q'}{k \ \Delta T_{av}} = \frac{Q}{q \ \tau} = \frac{Q'}{q}
\end{equation}

Type of the calculation formula for determining the average temperature head depends on the direction of motion of the coolant, which can move along the scheme: parallel flow, counter flow, cross current and mixed current. In parallel and counter flow, the average temperature difference between the heat carriers is determined by the equation:\\

\begin{equation} 
\Delta T_{av} = \frac{\Delta T_{b} - \Delta T_{s}}{ln \frac{\Delta T_{b}}{\Delta T_{s}}}
\end{equation}
where: \\
$\Delta T_{av}$ - average temperature difference, $K$; \\
$\Delta T_{b}$ - the biggest temperature difference between heat carriers, $K$; \\
$\Delta T_{s}$ - the smallest temperature difference between heat carriers, $K$.\\


The most common are the following partial indices of heat exchanger efficiency \cite{teplo2012}: 

$$\varepsilon = \frac{q}{q_{max}}$$
where $q$ -- rate of heat transfer from hot to cold coolant, $q_{max} = C_{min} (T^{in}_h - T^{in}_c)$. \par

\subsection{Criteria equations}
\
The coefficient of heat transfer from the wall to the liquid is determined from the criterial equations of convective heat transfer.
The heat transfer coefficient enters into the Nusselt criterion:

\begin{equation}
Nu = \frac{k \ l}{\lambda}
\end{equation}
where:\\
$Nu$ - Nusselt criterion;\\
$k$ - heat transfer coefficient, $W / (m2\cdot K)$;\\
$l$ - determining geometric dimension. $m$\\

\begin{equation}
l = D_h = \frac{4 \ A}{P}
\end{equation} 

\noindent $D_h$ - hydraulic (equivalent) diameter, $m$;\\
$A$ - the cross-sectional area of a liquid flow, $m^2$;\\
$P$ - full wetted perimeter of the flow section, $m$;\\
$\lambda$ - the thermal conductivity of a liquid, $W / (m\cdot K)$.\\

For a pipe of circular cross section, completely (without voids) filled with liquid, this formula takes the form:

$$D _ {h} = {\frac{4 {\frac {\pi D^{2}}{4}}} {\pi D}} = D $$
That is, for the circular section, the hydraulic diameter is equal to the geometric diameter. \\

In the case of forced motion of a liquid (gas) in tubes and channels, the specific form of the criterial equation for determining the Nusselt criterion depends on the fluid (gas) flow regime.

The motion regime is characterized by the value of the Reynolds criterion: \\

\begin{equation}
Re = \frac{w \ l \ \rho}{\mu} = \frac{w \ l}{\nu}
\end{equation}
where: \\
$Re$ - is the Reynolds criterion;\\
$w$ - fluid velocity, $m / s$;\\
$l$ - defining geometrical size, $m$;\\
$\rho$ - is the density of the liquid, $kg / m^3$;\\
$\mu$ - is the dynamic coefficient of viscosity of the liquid, $Pa \cdot s$;\\
$\nu$ - is the kinematic coefficient of viscosity of the liquid, $m^2 / s$.\\

In the case of a developed turbulent regime of fluid motion in straight pipes and channels (Re $>$ 10000), the Nusselt criterion is determined by the equation: \\

\begin{equation}
Nu = 0.021 \ \varepsilon_1 \ Re^{0.8} \ Pr^{0.43} \ \left( \frac{Pr}{Pr_{w}} \right) ^{0.25}
\end{equation}

\noindent where: \\
$\varepsilon$ - is the correction factor taking into account the influence on the heat transfer coefficient of the ratio of the length of the heat exchanger tubes to their diameter; At $\frac{l}{d} >= 50 \ \ \  \varepsilon = 1 $;\\
$Pr$ - is Prandtl's criterion, calculated for a liquid at an average temperature:\\

\begin{equation}
Pr = \frac{c \ \mu}{\lambda}
\end{equation}

\noindent where: \\
$c$ - specific heat of the liquid, $J / (kg \cdot K)$;\\
$\mu$ - is the dynamic coefficient of viscosity of the liquid, $Pa \cdot s$;\\
$\lambda$ - is the thermal conductivity of the liquid, $W / (m \cdot K)$;\\
$Pr_{w}$ - Prandtl's criterion, calculated for a liquid at the temperature of the wall.\\

In the criterial equations of convective heat transfer there is a relation $\left( \frac{Pr}{Pr_{w}} \right) ^{0.25}$, which takes into account the direction of the heat flux. For droplet liquids, with increasing temperature, the value of the Prandtl criterion decreases, therefore, when the liquids H are heated $\frac{Pr}{Pr_{w}} > 1$, and when the liquids cool $\frac{Pr}{Pr_{w}} < 1$. When designing heat exchangers in calculating the heat transfer coefficient for heated liquids, it is allowed to assume $\frac{Pr}{Pr_{w}} = 1$, and for cooling liquids one can take the average value $\left( \frac{Pr}{Pr_{w}} \right) ^{0.25} = 0.93.$ \par

For gases, the equation for calculating the Nusselt criterion is simplified, since $\frac{Pr}{Pr_{w}} = 1$, and the Prandtl criterion depends on the atomicness of the gases: \\

\begin{tabular}{l l}
monatomic     gases & $Pr=0.67$\\
diatomic gases & $Pr=0.72$\\
four and polyatomic gases & $Pr=1.0$ \\
\end{tabular} 
\bigskip

\noindent For example, for air the criterial equation takes the form:

\begin{equation}
Nu = 0.018 \ \varepsilon_1 \ Re^{0.8} 
\end{equation} 

Under the laminar flow regime of liquid in straight pipes and channels (Re $<$ 2300), the Nusselt number can be determined from the following equations:

\begin{equation}
Nu = 0.17 \ \varepsilon_1 \ Re^{0.33} \ Pr^{0.43} \ Gr^{0.1} \ \left( \frac{Pr}{Pr_{w}} \right) ^{0.25} 
\end{equation}

Or

\begin{equation}
Nu = 1.4 \ \left( Re \frac{d}{L} \right)^{0.4} \ Pr^{0.33} \  \left( \frac{Pr}{Pr_{w}} \right) ^{0.25}
\end{equation}

And also

\begin{equation}
Nu = 1.55 \ \left( Pe \frac{d}{L}\right) ^{0.33} \left( \frac{\mu}{\mu_w} \right)^{0.14} = 1.55 \ \left( Re \ Pr \frac{d}{L}\right) ^{0.33} \left( \frac{\mu}{\mu_w} \right)^{0.14} 
\end{equation}

\noindent where: \\
$L$ - is the length of the pipe, $m$;\\
$D$- Pipe diameter, $m$; \\
$ Pe = Re / Pr$ - the Peclet criterion;\\
$Gr$ - is Grashof's criterion:

\begin{equation}
Gr = \frac{g \ l^3 \ \beta \ \left( T_{wall} - T_{av} \right)}{\nu^2} 
\end{equation}

\begin{description}
\item[$\beta$] is the volume expansion coefficient of the liquid (gas), $K^{-1}$.
\end{description}

With heat transfer in the transient mode of fluid motion 2300 $<$ Re $<$ 10000 Nusselt criterion is determined from the graphical dependence. The graph of the dependence is shown in Figure 1. 
$$ \frac{Nu}{Pr^{0.43} \ \left( \frac{Pr}{Pr_{w}} \right) ^{0.25}} = f(Re)$$

\begin{figure}
\includegraphics[width=0.7\textwidth]{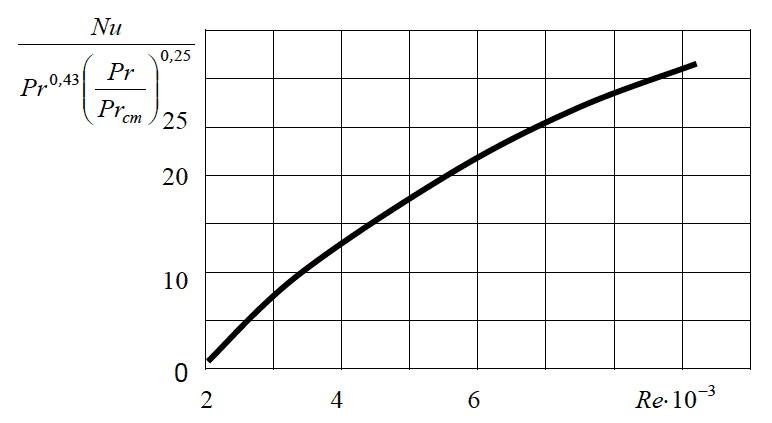}
\centering
\caption[Short figure name.]{Dependence of the Nusselt criterion on the Reynolds criterion under transient conditions. \cite{b0023}
\label{fig:myInlineFigure}}
\end{figure}

And also by the approximate equation

\begin{equation}
Nu = 0.008 \ Re^{0.9}  \ Pr^{0.43} 
\end{equation}
\\

\subsection{Constructive and hydrodynamic calculation of heat exchanger}
The purpose of constructive calculation is to determine the main dimensions of the heat exchanger. The task is to determine the number of pipes, their arrangement, the diameter of the apparatus, the number of strokes in the pipe and intertubular spaces and the dimensions of the nozzles. \par
For shell-and-tube heat exchanger, having the greatest distribution in industry, by the heat exchange surface the number of pipes is determined, their placement in the tube grid, the diameter of the body of the equipment, the number of strokes in pipe's and intertubular space and the dimensions of the inlet and outlet tubes. \\ 

Determination of the number of pipes is based on the heat exchange surface F calculated in chapter 2: 

\begin{equation}
n = \frac{F}{\pi \ D_n \ l}
\end{equation}

\noindent where: \\
$n$ - the number of pipes; \\
$D_n$ - nominal pipe diameter, $m$; \\
$l$ - the length of the pipes, $m$.\\

In the case of an incompressible fluid the frictional resistance at the motion of the coolant in the channels is determined \cite{Isachenko}:

\begin{equation}
\Delta P_f = \xi \frac{l}{d} \frac{\rho V^2}{2}
\end{equation}
where:\\
$l$ - total length of the channel;\\
$d$ - pipe diameter or equivalent hydraulic diameter of the channel;\\
$\xi$ - is the coefficient of friction resistance;\\
$\rho$ - the media density;\\
$V$ - the media velocity.\\

To calculate all parameters of a heat exchanger an Excel spreadsheet was created. It is very useful for an experimental part to understand how the heat exchangers looks like.

\section{Modelling}

\subsection{Model mesh and computer parameters} 

To fulfill quality of computational model -- sufficiency of equations to solve the problem, unchanged error on convergence plot and enough computational power of a computer, the following mesh was  set up: \\ \linebreak
For inner area: \\
- maximum element size: 24.3 mm,\\
- minimum element size: 6.07 mm,\\
- maximum element growth rate: 1.3,\\
- curvature factor: 0.9,\\
- resolution of narrow regions: 0.4.\\

\noindent For boundaries: \\
- maximum element size: 15.8 mm,\\
- minimum element size: 4.85 mm,\\
- maximum element growth rate: 1.25,\\
- curvature factor: 0.8,\\
- resolution of narrow regions: 0.5.\\

All models were calculated in $COMSOL^{ \scriptsize \textregistered} \ 5.2a$ with this mesh, if not stated otherwise, and on a computer with the following parameters: processor: Intel Core i7 4960X 3.60 GHz; 64 GB RAM; SSD Kingston HyperX 256GB; graphics card: Nvidia Quadro k600. Time of computing for  models below is up to 20 hours (cases with twisted tubes). On graphs middle line of the central tube $( (0,0,0);(0,0,1500) )$ was taken as a data for all models. All material's constants and variables were taken from the build-in tables.

\subsection{Heat exchanger with 19 tubes in full size}
\label{S:2}

Full sized heat exchanger was created in two cases: with plain tubes inside a shell and with twisted tubes. Air flows through tubes and water goes between tubes and shell. Plain tube has a diameter 12 mm ($D_h$ is 12 mm), shell is 1500 mm in length and 100 mm in diameter, twisted tube (Figure 2) has a relation in ellipse's axes 6.6 x 5.28 mm ( $D_h$ is 11.70 mm) and twisted 8 times along the length. Thickness of shell is 2 mm a tube's one is 1 mm. Water inlet and outlet has a diameter 40 mm and is located 150 mm from the edge. 


\begin{figure} [H]
\includegraphics[width=0.8\textwidth]{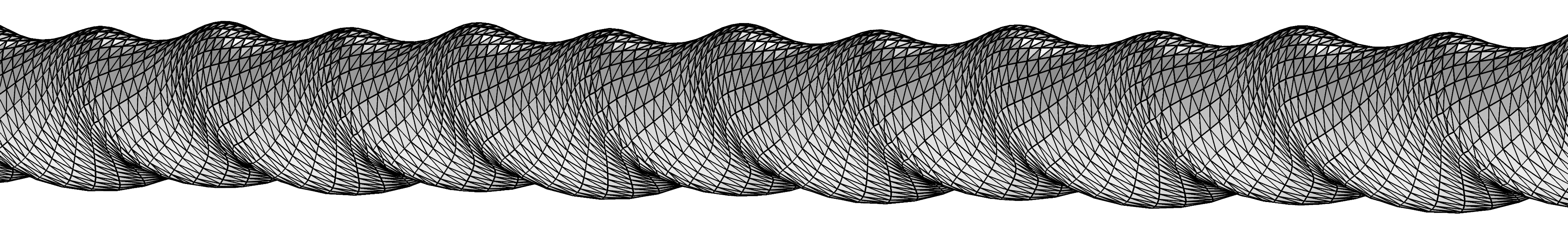}
\centering
\caption[Short figure name.]{Meshed twisted tube from the side.
\label{fig:myInlineFigure}}
\end{figure}

Temperature and velocity for air was set for each tube separately. Inlet air's temperature is $500 ^{\circ}\mathrm{C}\ (773.15 \ K)$ and velocity is 23 m/s; inlet temperature of water is $10 ^{\circ}\mathrm{C}\ (283.15 \ K)$ and velocity is 1.05 m/s. These inputs numbers was equal to flowing water and maximum temperature of the heater at this temperature and a flow rate of 1500 L/min. 
For air and water turbulent flow $k-\varepsilon$ physics model was chosen as Reynolds number for air in mixed flow is about 2000. 
The results of the computation are presented on the Figures 3--4. \bigskip

It can be seen that in twisted case air cools faster that in plain one and the most intensive heat transfer occurs at first 1/3 part of a heat exchanger. Also two cases with twisted tubes of a hydraulic diameter 10.63 mm and 13.29 mm were computed. All inlet parameter are the same, the only change is tube size. Middle line of the central tube $( (0,0,0);(0,0,1500) )$ was taken as a data and the results of all 4 cases can be seen on Figure 5.\\

 \begin{figure}[H]

\subfloat[Velocity \label{subfig-1:dummy}]{\includegraphics[width=0.5\textwidth]{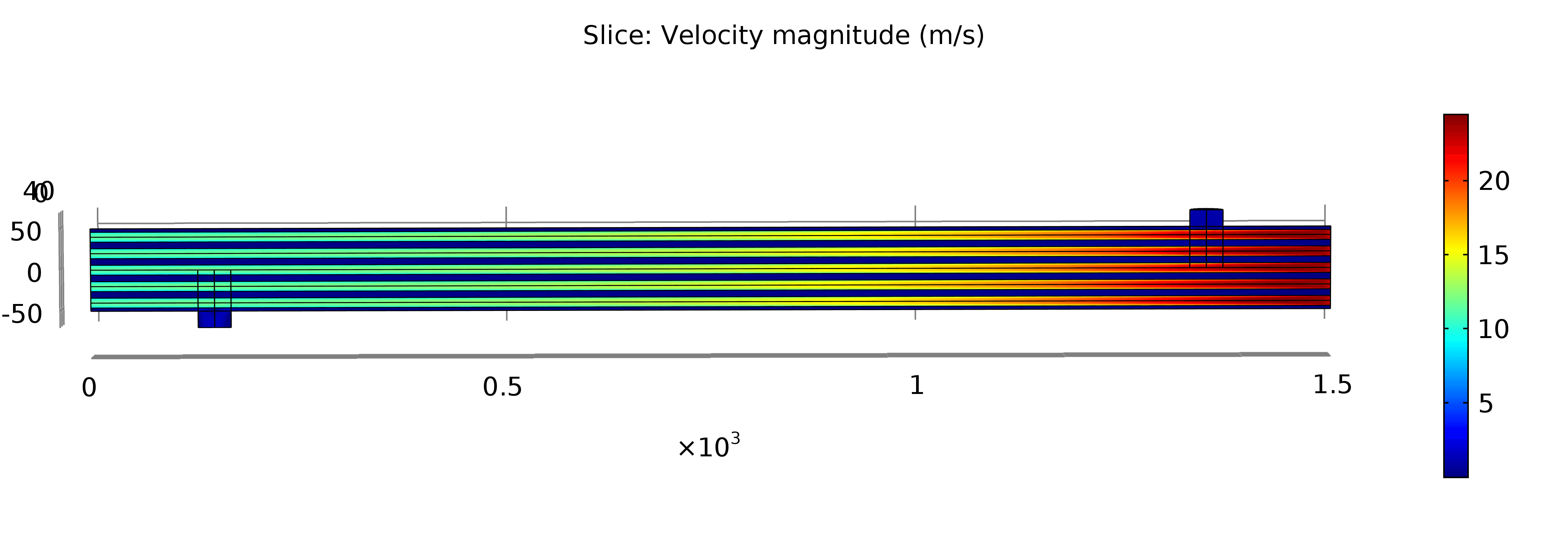}}
\subfloat[ Temperature \label{subfig-2:dummy} ]{\includegraphics[width=0.5\textwidth]{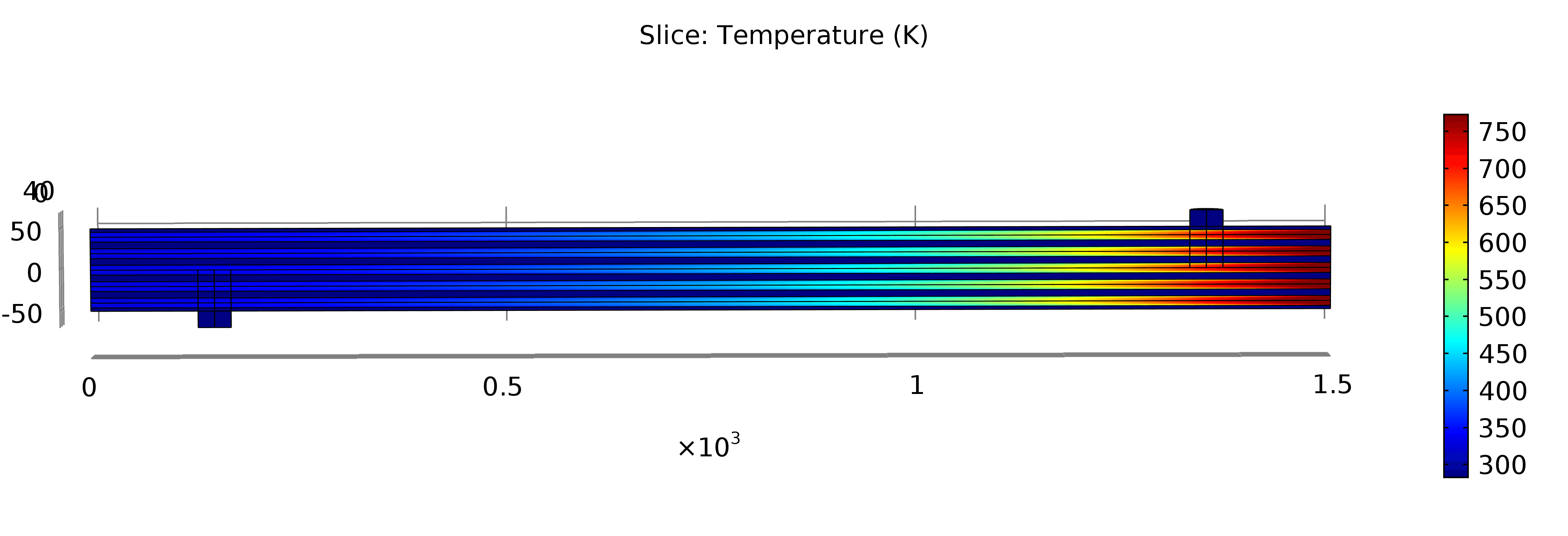}}
\hfill
\centering
\caption{Results for plain tubes, $D_h$ = 12 mm.}
\label{fig:dummy}

\end{figure}

 \begin{figure}[H]

\subfloat[Velocity \label{subfig-1:dummy}] {\includegraphics[width=0.5\textwidth]{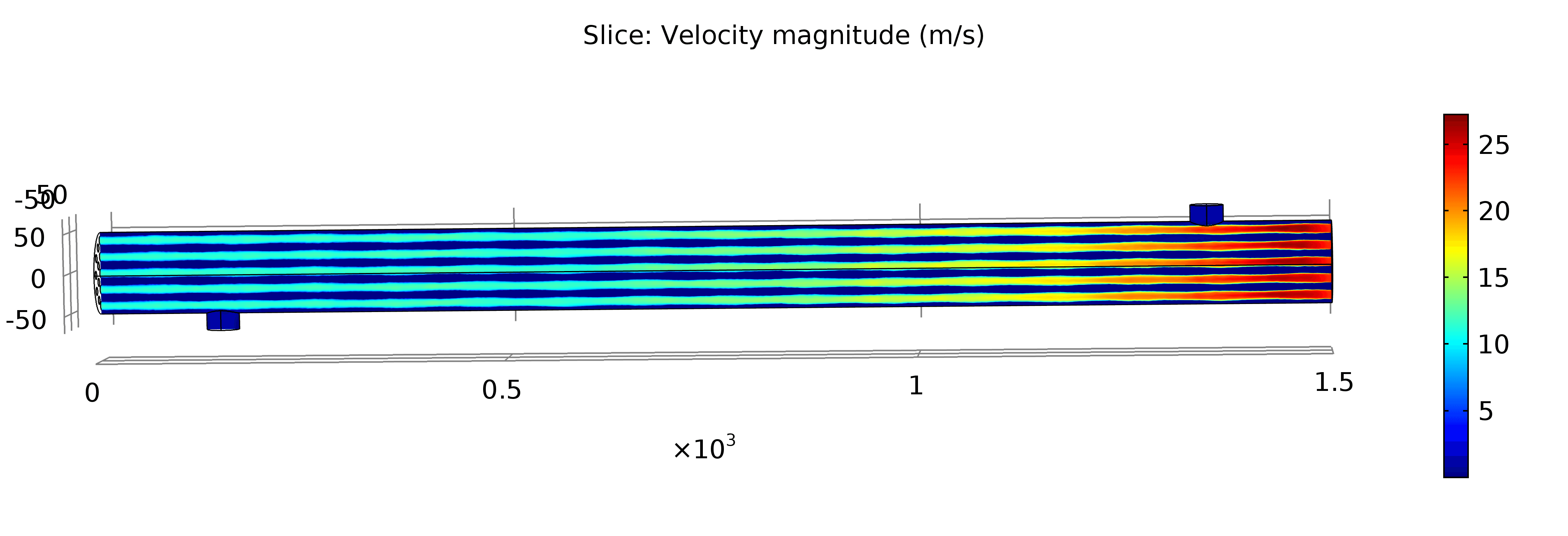}}
\subfloat[Temperature \label{subfig-2:dummy} ]{\includegraphics[width=0.5\textwidth]{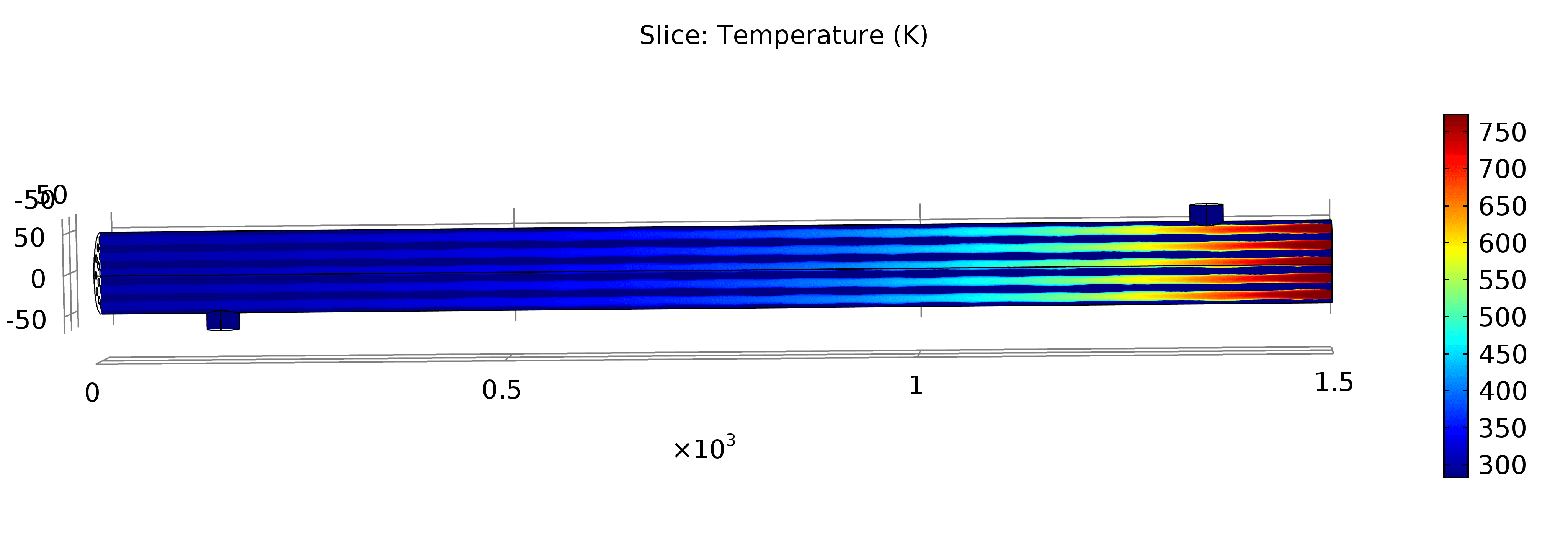}}
\hfill
\centering
\caption{Results for twisted tubes, $D_h$ = 11.70 mm.} 
\label{fig:dummy}

\end{figure}

\bigskip

 \begin{figure}

\subfloat[Velocity  \label{subfig-1:dummy}] {\includegraphics[width=\textwidth]{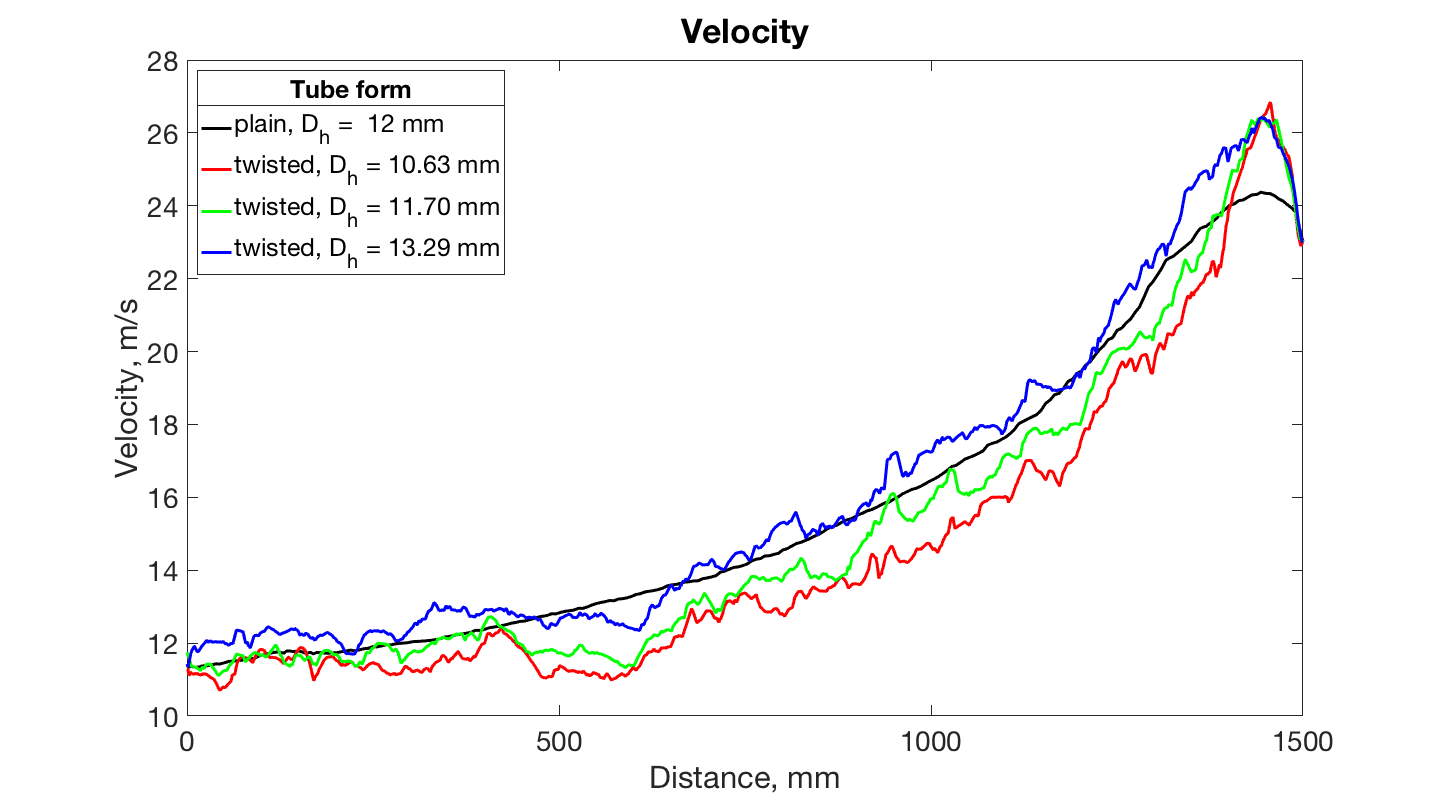}}
\hfill
\subfloat[Temperature  \label{subfig-2:dummy} ]{\includegraphics[width=\textwidth]{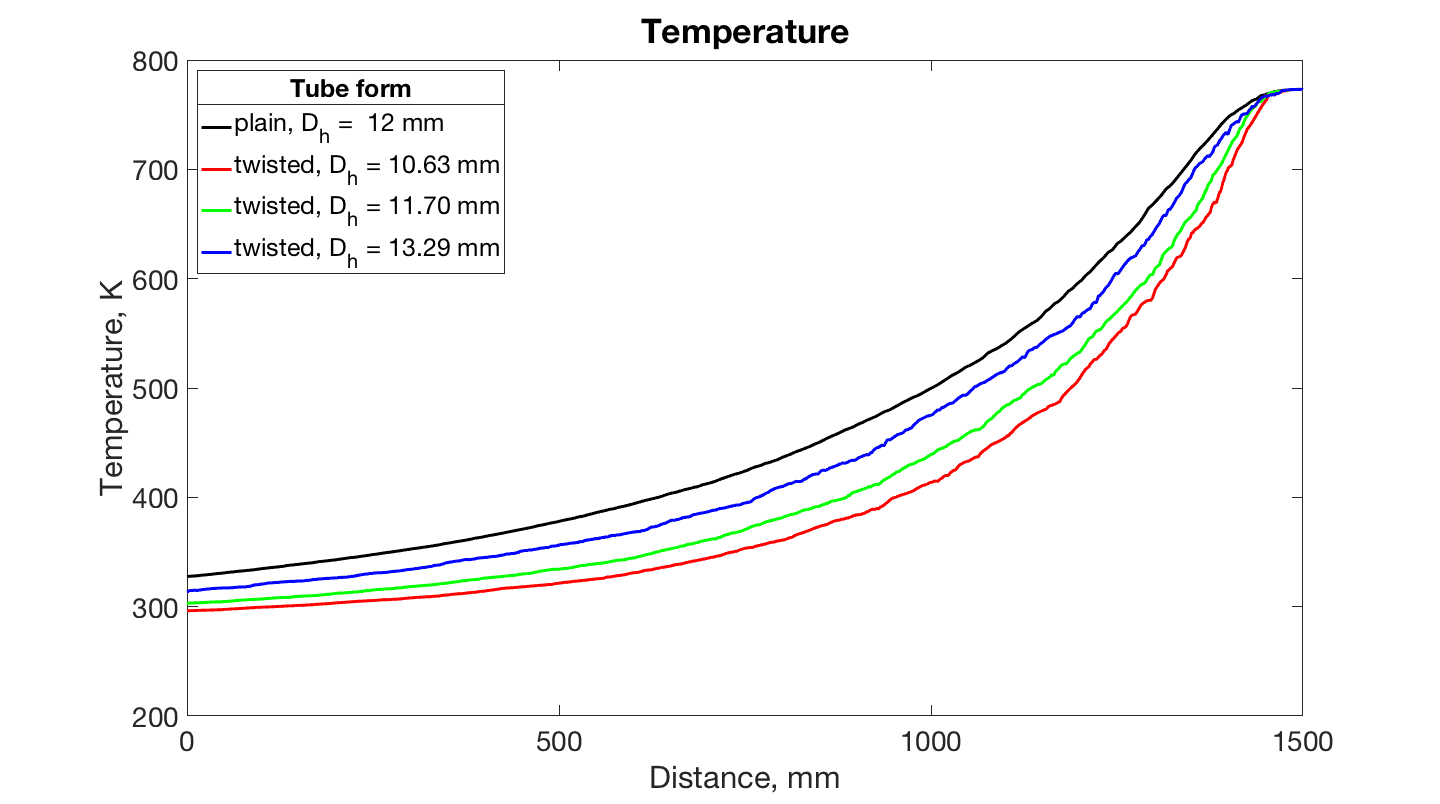}}
\hfill
\centering
\caption{Results for all four cases at section 3.2.}
\label{fig:dummy}

\end{figure}

On the temperature figure an effect of better efficiency of twisted tubes can be seen. Final temperature in each cases are: plain, $D_h$ = 12 mm -- 327.637 K; twisted, $D_h$ = 10.63 mm -- 296.266 K;  twisted, $D_h$ = 11.70 mm -- 303.144 K; twisted, $D_h$ = 13.29 mm -- 313.541 K. The result are according a theory, as the smaller diameter of the tube, the greater heat transfer coefficient. Red plot is less in temperature for about 10\% than at plain case. Moreover, the tube size in this case is less, that means less material for the whole system and cheaper system in production. It is highly important in industry. At a velocity graph it can be seen growth of all twisted cases than plain one at the beginning and next decreasing of a velocity. This effect is connected with tube geometry - periodically changes in tube's height and mixing of a flow along the whole tube.\par

Advantages of twisted tube are: increased heat transfer (40\% higher tubeside heat transfer coefficient), smaller exchangers or fewer shells, elimination of flow-induced vibration, and reduced fouling. When used as retrofit bundles or exchangers, Twisted Tube tubings also offer increased capacity, lower installed costs, lower pressure drop, and extended run time between cleanings\cite{handbook}. Nowadays several companies are produced heat exchanger with twisted tubes: Koch Heat transfer company \cite{koch}, Eddify \cite{eddi}, Trinvalco \cite{trin} and ABI \cite{abi}. \par

Numerical results in \cite{1-s2} show that the axial stiffness of twisted tubes decreases with increasing the twist ratio or decreasing the lead. With the stiffness reduction twisted tube heat exchanger could be more accurately designed. It is hard to find out research paper where twisted tube model is analysed. In \cite{p1}, \cite{p2}, \cite{p3} twisted-tape element is studied. \\

\textbf{Summary} \medskip

Four models were developed with different geometry and size of tubes. It was shown that twisted tube has higher heat transfer than plain tube. Decreasing of a diameter of the tube inside a shell leads to increasing of thermal efficiency, i.e. lower temperature at the end of a tube at the same initial parameters. \\

\subsection{Heat exchanger with common inlet \& outlet}


In previous section heat exchanger with 19 tubes inside a shell in full size was computed. But inlet velocity and temperature were set equal in each tube. In real case inlet water and air flows are not uniform alone a vertical slice. As it is known flow inside a tube has a parabolic form, where velocity of the flow in a center is higher than near edges because of a viscosity. Front flange is an obstacle for air, because some air flow pass further through a tube and some bump into the metal wall, that cause turbulence near a heat exchanger inlet and velocity jump inside a tube. Common inlet and outlet should be designed and will be considered in this part. The same two models - with plain and twisted tubes were developed. Length of the inlet and outlet part is 100 mm. All parameters remains the same as in previous section: inlet air's temperature is $500 ^{\circ}\mathrm{C}\ (773.15 \ K)$ and velocity is 23 \ m/s; inlet temperature of water is $10 ^{\circ}\mathrm{C}\ (283.15 \ K)$ and velocity is 1.05 m/s. Two models with plain and twisted tubes are presented on the Figure 6 and Figure 7. and all results are on Figure 8. Green dashed line shows locations of heat exchanger inlet and outlet. \\

 \begin{figure}[H]

\subfloat[Velocity \label{subfig-1:dummy}] {\includegraphics[width=0.5\textwidth]{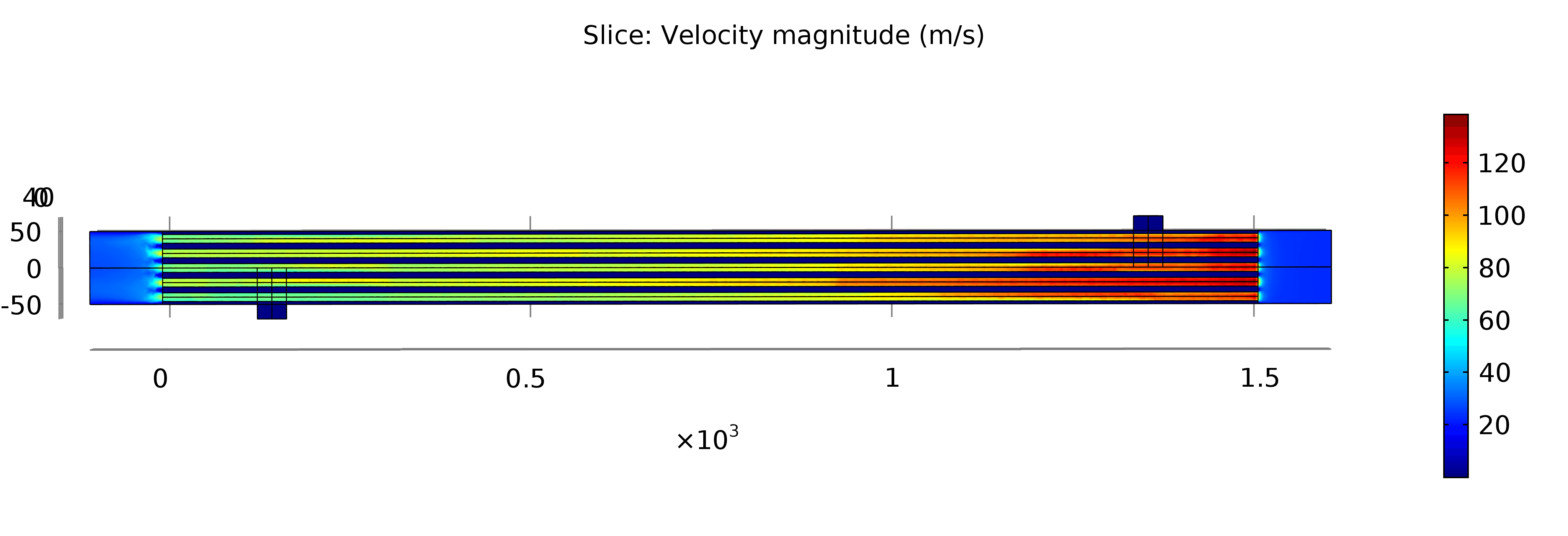}}
\subfloat[Temperature \label{subfig-2:dummy} ]{\includegraphics[width=0.5\textwidth]{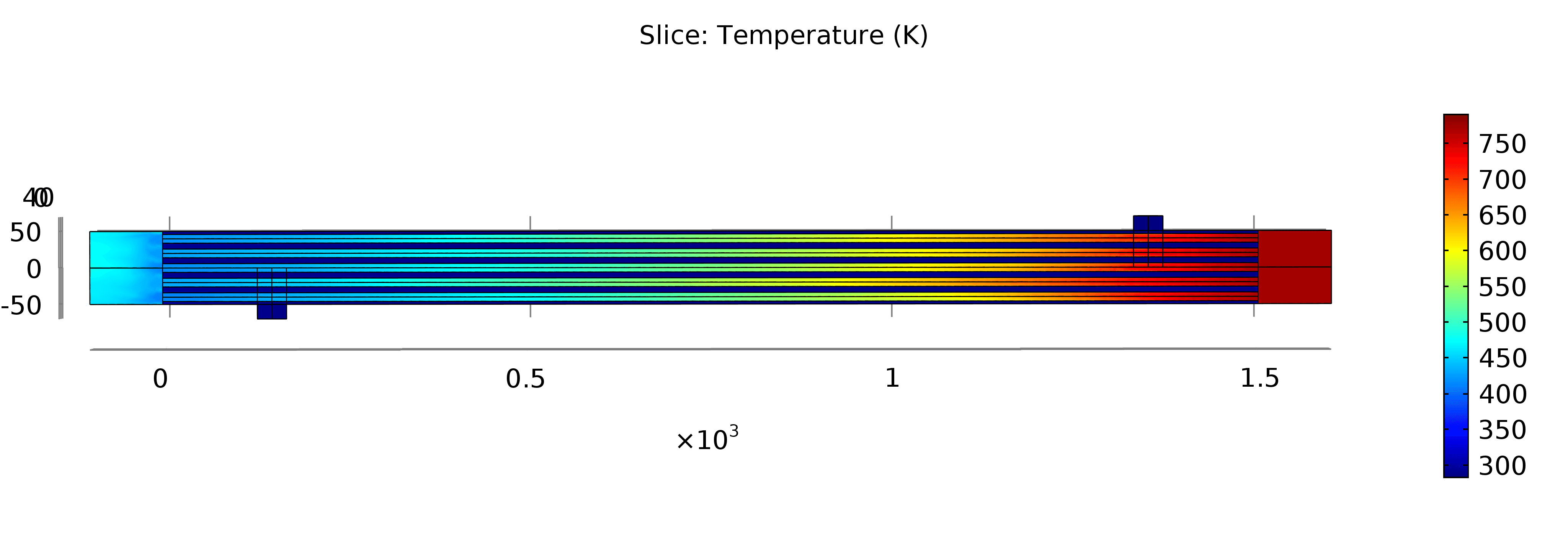}}
\hfill
\centering
\caption{Results for plain tubes, $D_h$ = 12 mm.}
\label{fig:dummy}

\end{figure}

 \begin{figure}[H]

\subfloat[Velocity  \label{subfig-1:dummy}] {\includegraphics[width=0.5\textwidth]{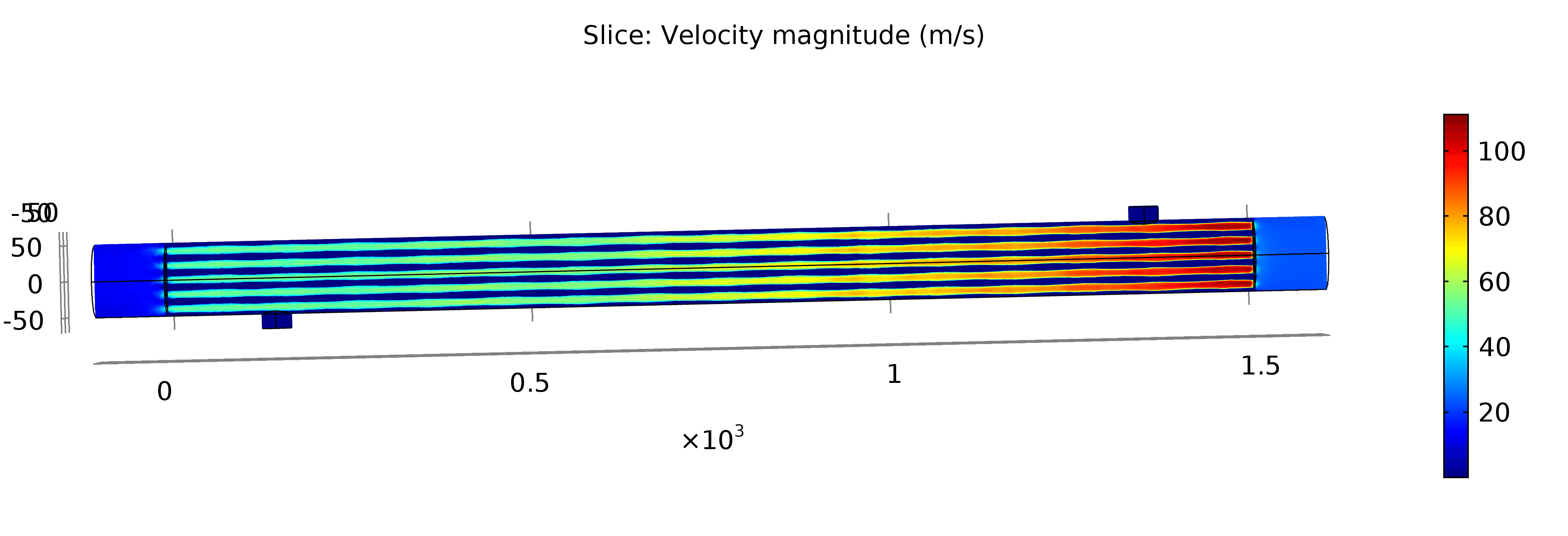}}
\subfloat[Temperature  \label{subfig-2:dummy} ]{\includegraphics[width=0.5\textwidth]{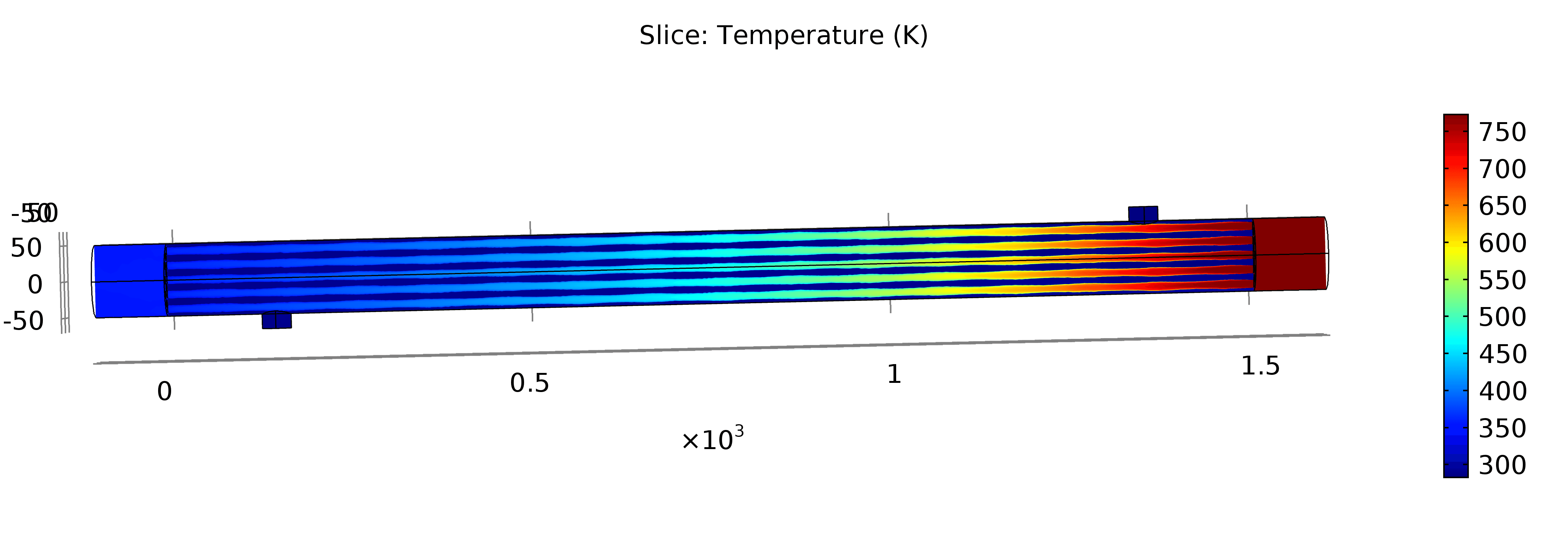}}
\hfill
\centering
\caption{Results for twisted tubes, $D_h$ = 11.70 mm.}
\label{fig:dummy}

\end{figure}

 \begin{figure}

\subfloat[Velocity \label{subfig-1:dummy}] {\includegraphics[width=0.95\textwidth]{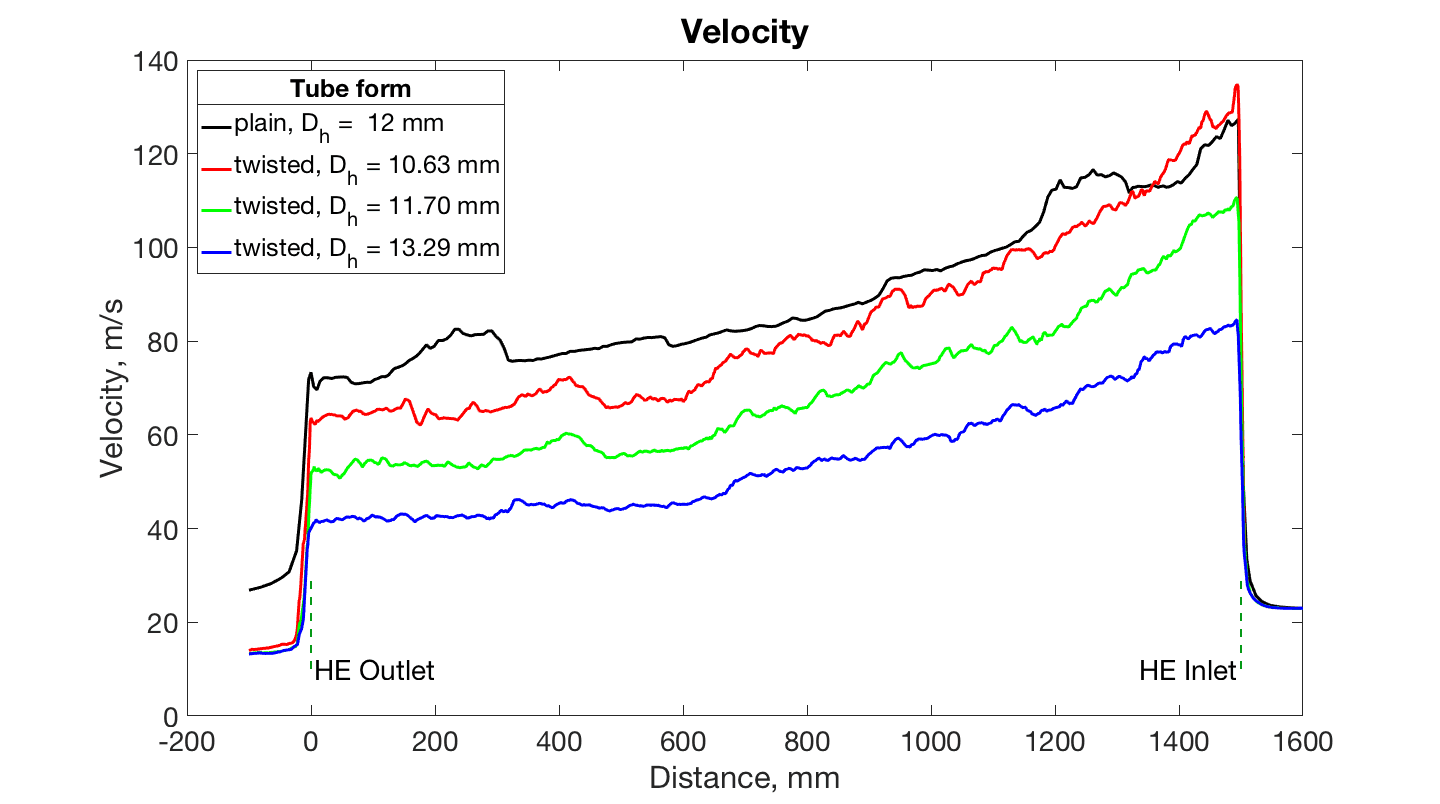}}
\hfill
\subfloat[Temperature \label{subfig-2:dummy} ]{\includegraphics[width=0.95\textwidth]{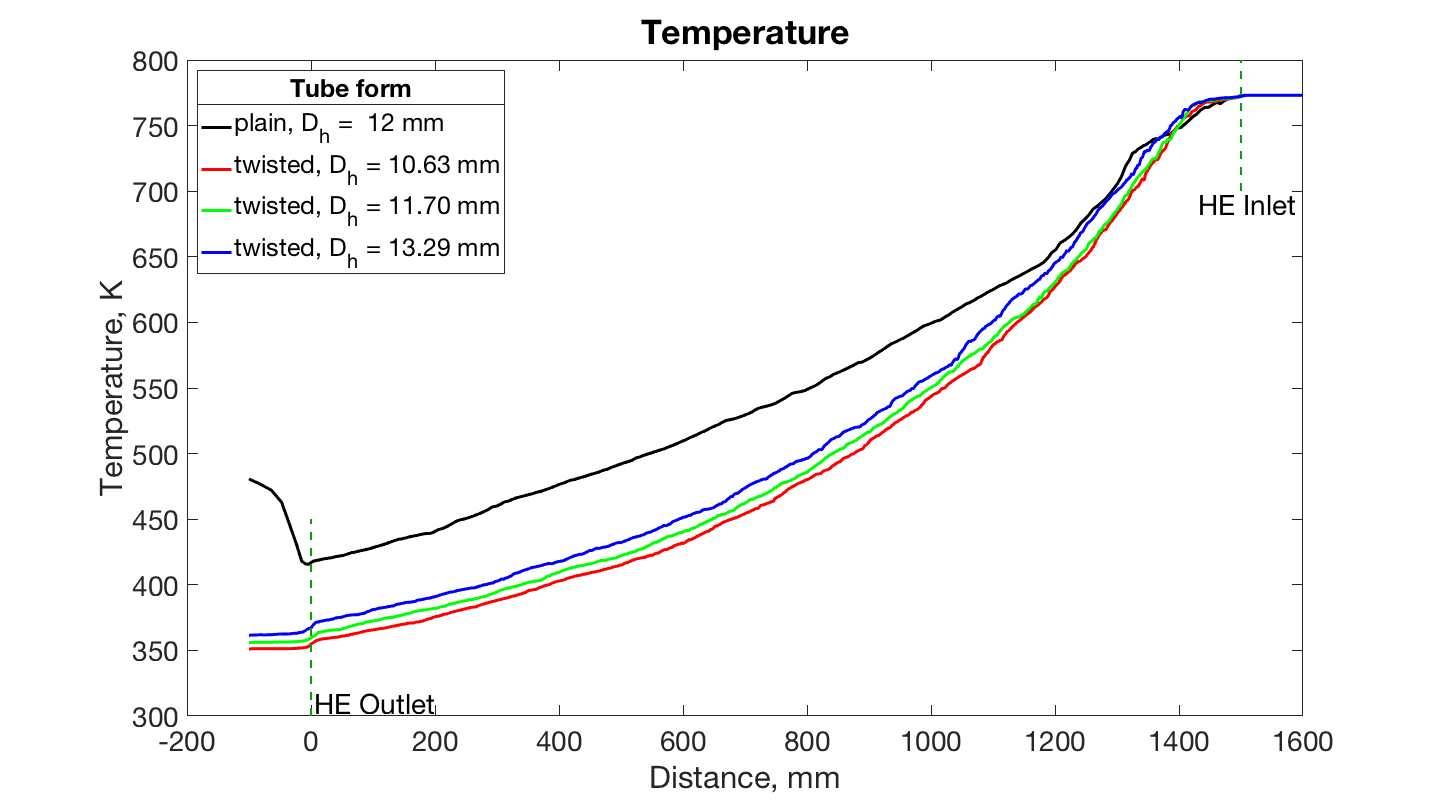}}
\hfill
\centering
\caption{Results for all four cases at section 3.3.}
\label{fig:dummy}

\end{figure}

Significant difference in temperature can be seen on a results' figure. Final temperature in each cases are: plain, $D_h$ = 12 mm  - 480.696 K; twisted, $D_h$ = 10.63 mm - 351.169 K;  twisted, $D_h$ = 11.70 mm - 355.909 K; twisted, $D_h$ = 13.29 mm - 361.530 K. Here gain of red case over the black one is 27\% in temperature. This is an extremely efficiency increase of a heat exchanger. Flow velocity increased up to 130 m/s at the beginning of a heat exchanger. Final velocity for plain and twisted cases are 27 m/s and 13 m/s respectively. Now whole length of a HE is used to cool air. For twisted cases velocity growth proportionally to a decreasing of a hydraulic diameter. It can be seen on the temperature plot, that in first 300 mm air cools approximately equally for all cases and then plain case slow down decreasing and have higher temperature at the end of whole system. Rise of temperature in plain case after the heat exchanger outlet is due to higher outlet temperature of other tubes that mixed in this area. For twisted case opposite situation can be seen. \\

\textbf{Summary} \medskip

Four models were developed with different geometry and size of tubes with common inlet and outlet for air flow. It was shown that twisted tube has higher heat transfer than plain tube and temperature difference is significant. Change in hydraulic diameter have approximately the same dependence as in section 3.2. \\

\subsection{Baffles inside a heat exchanger} 

Baffles inside a shell increase intensity of heat transfer. There can be different approaches how to set them. First way is to put baffles equally alone the length of a heat exchanger. Half circle was taken in the first model and put each 250 mm rotating on $180^{\circ}$. Next approach is to set baffles non uniformly. As the biggest drop in temperature was observed at first 1/3 of a heat exchanger, then it should be suitable to put more baffles near the entrance of a water flow to increase intensity of mixing. It is done in the second model. One more case is to take baffle as 3/4 circular segment of a circle. As was written in theory part it is an optimal cut for a baffle and  should increase heat transfer. The last modification of baffle is made as two circle sectors with angle $90^{\circ}$ opposite to each other (Figure 9). It should mix water flow and increase heat transfer. This construction, looks like "pizza", was set each 50mm between 250mm and 500mm and after that each 250mm and is rotated on $180^{\circ}$ every step. All models (temperature and velocity slices) are shown on the Figures 10--13 and the results of all five cases are presented on the Figure 14.

\begin{figure}[H]
\includegraphics[height=4cm, width=0.95\textwidth]{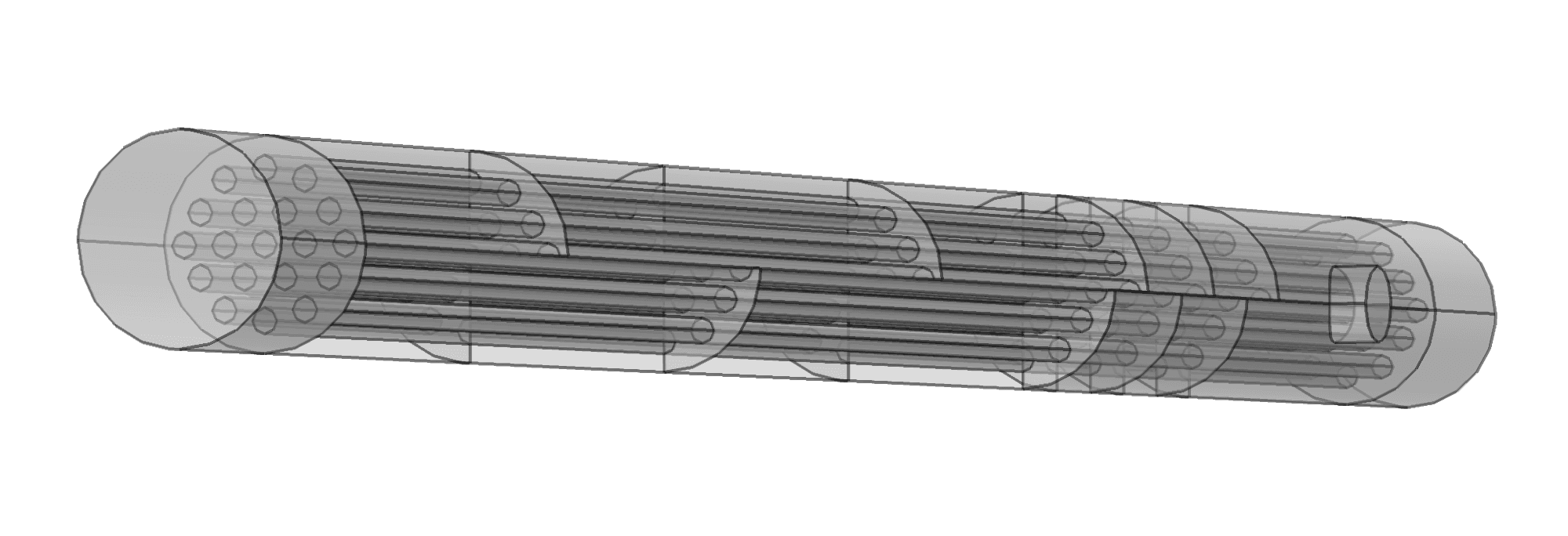}
\centering
\caption[Short figure name.]{Baffle as two opposite to each other circle sectors with angle $90^{\circ}$ .
\label{fig:myInlineFigure}}
\end{figure}

\begin{figure}[H]

\subfloat[Velocity \label{subfig-1:dummy}] {\includegraphics[width=0.5\textwidth]{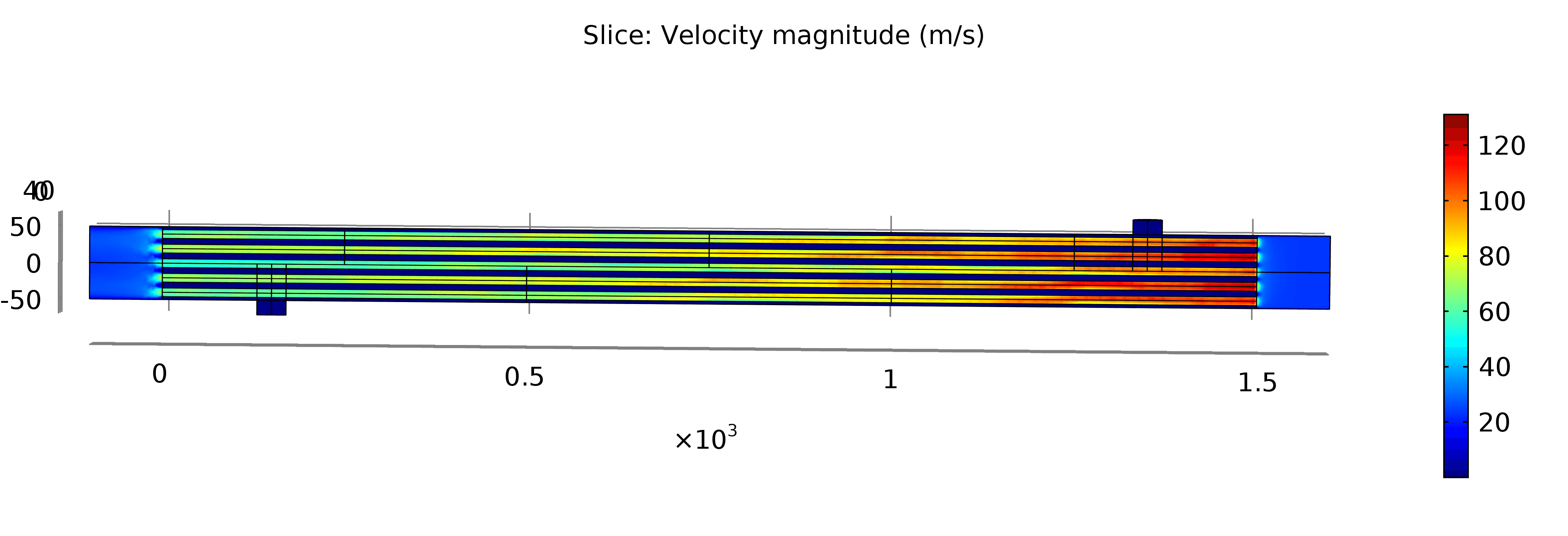}}
\subfloat[Temperature \label{subfig-2:dummy} ]{\includegraphics[width=0.5\textwidth]{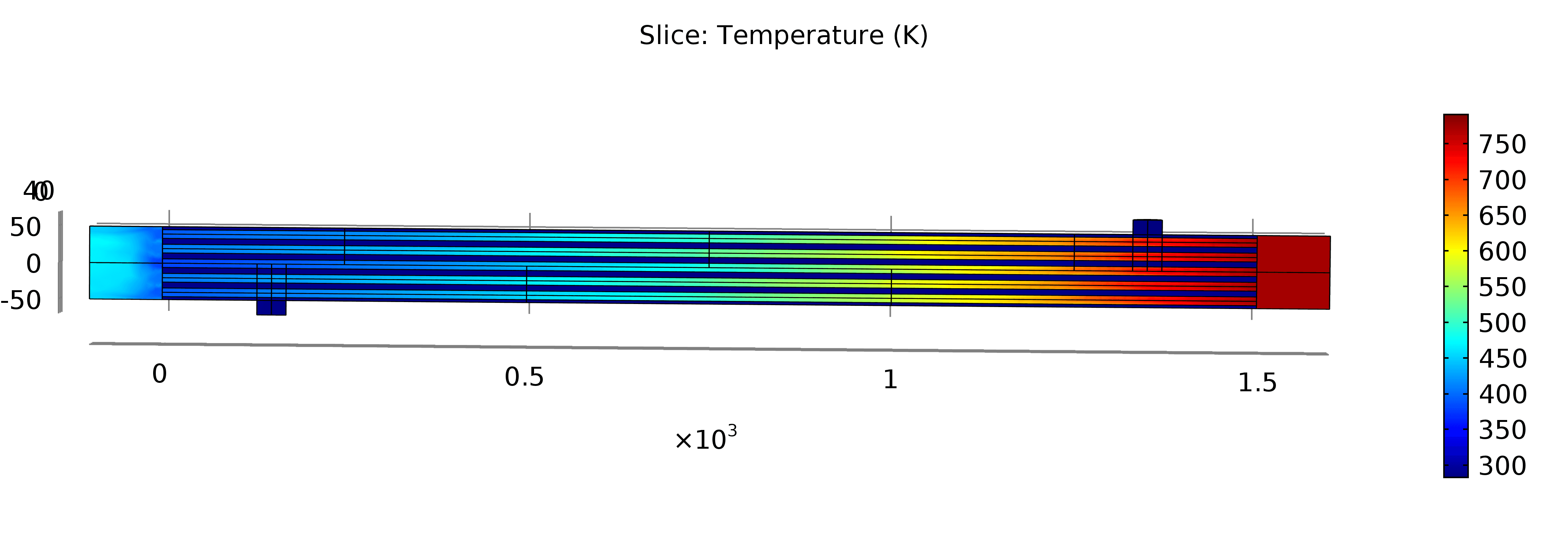}}
\hfill
\centering
\caption{Results for plain tubes with half circle baffles.}
\label{fig:dummy}

\end{figure}

\begin{figure}[H]

\subfloat[Velocity \label{subfig-1:dummy}] {\includegraphics[width=0.5\textwidth]{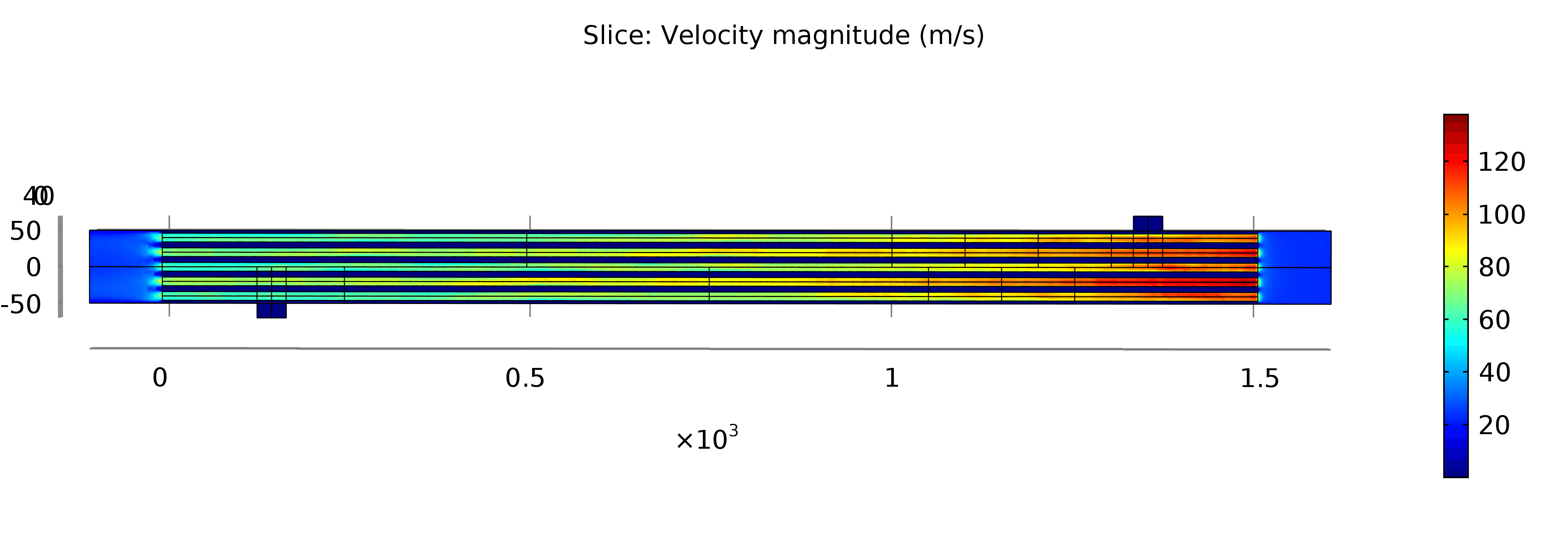}}
\subfloat[Temperature \label{subfig-2:dummy} ]{\includegraphics[width=0.5\textwidth]{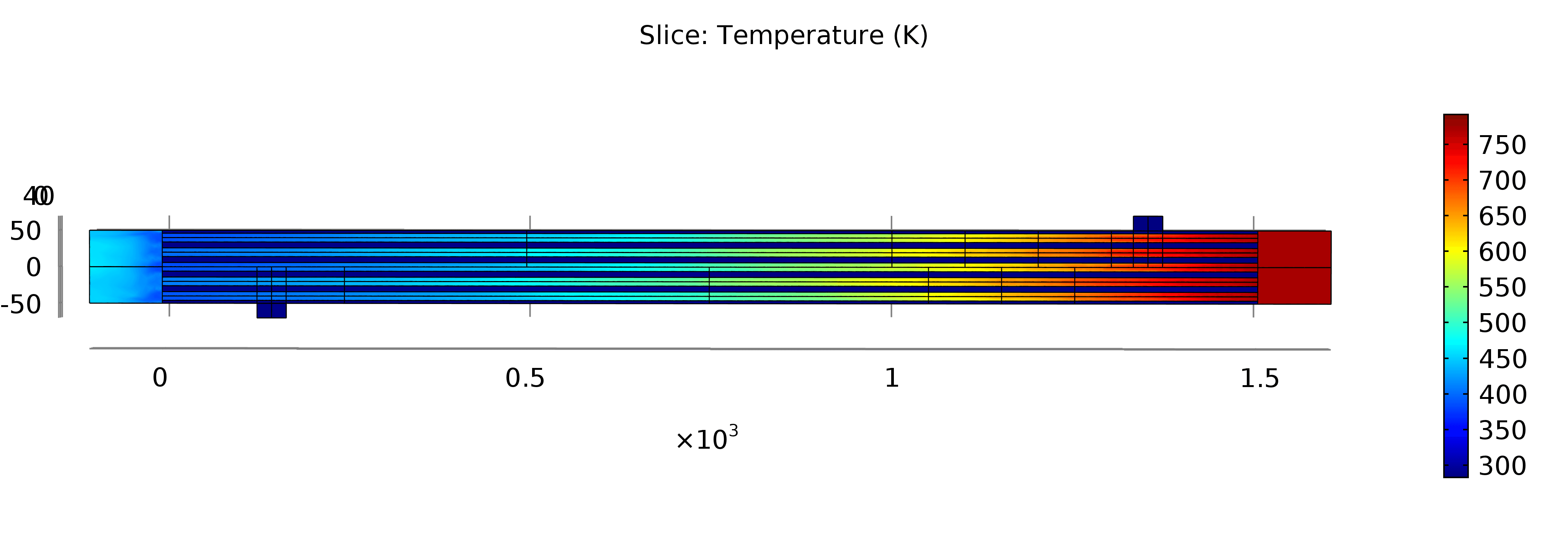}}
\hfill
\centering
\caption{Results for plain tubes with more baffles near water inlet.}
\label{fig:dummy}

\end{figure}

\begin{figure}[H]

\subfloat[Velocity \label{subfig-1:dummy}] {\includegraphics[width=0.5\textwidth]{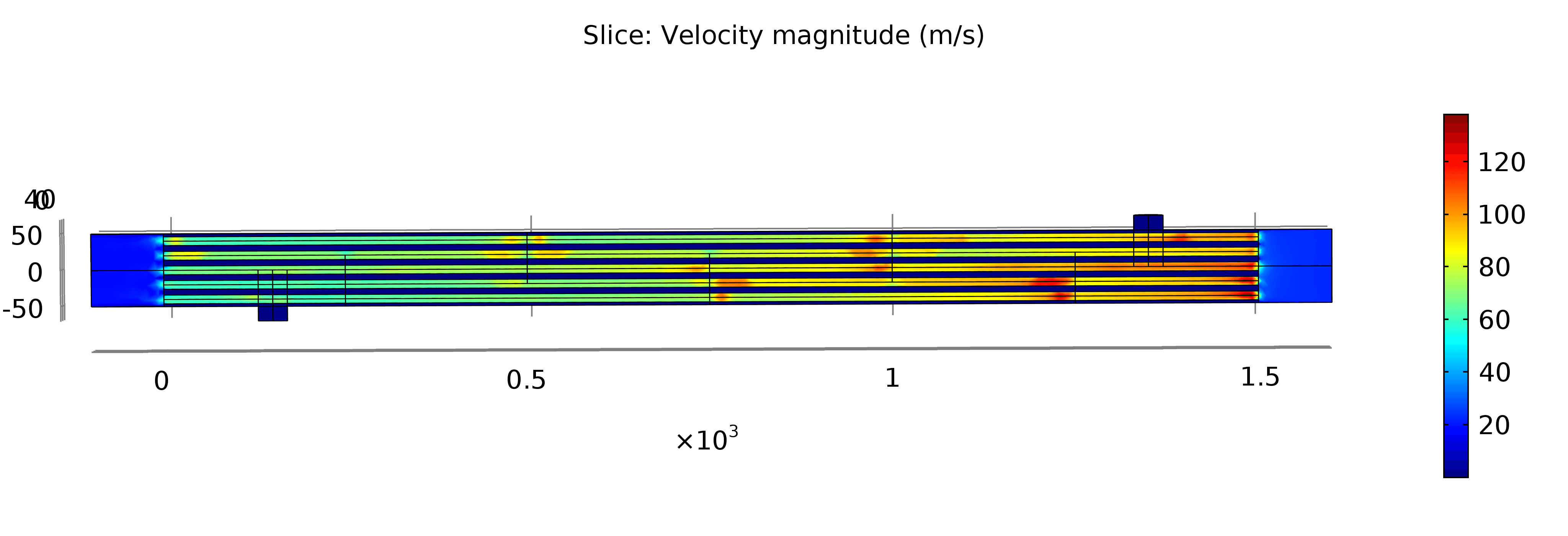}}
\subfloat[Temperature \label{subfig-2:dummy} ]{\includegraphics[width=0.5\textwidth]{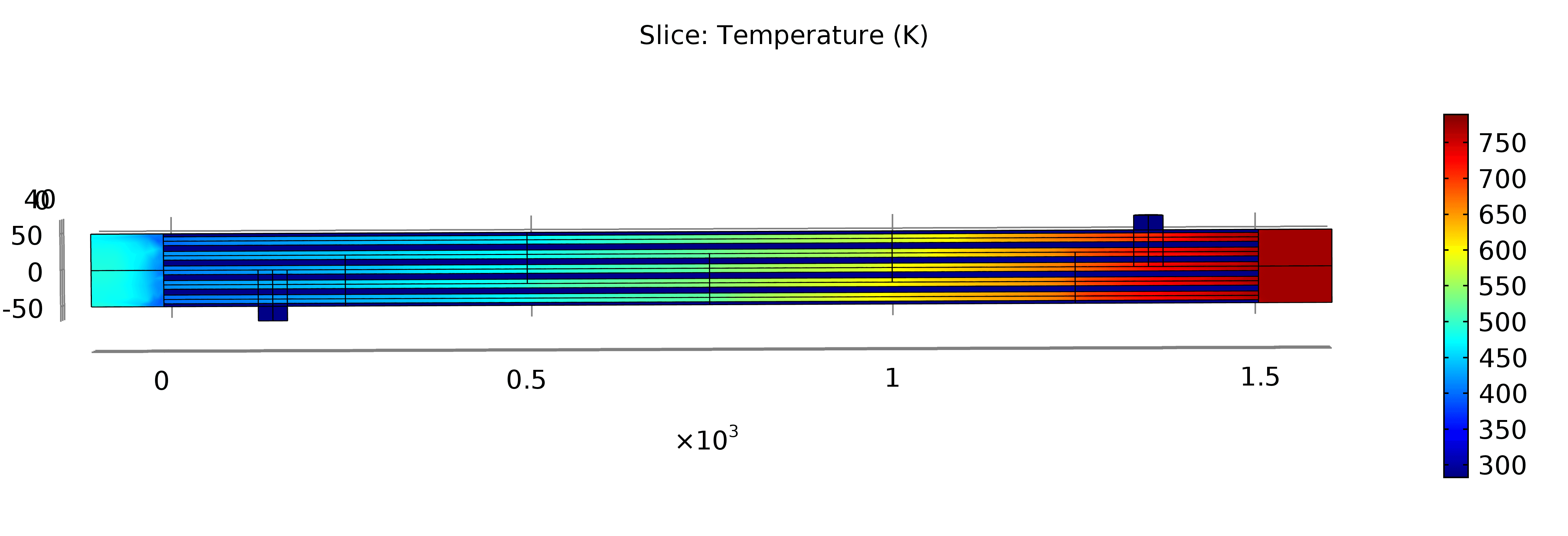}}
\hfill
\centering
\caption{Results for plain tubes with 3/4 circular segment baffles.}
\label{fig:dummy}

\end{figure}

 \begin{figure}[H]

\subfloat[Velocity \label{subfig-1:dummy}] {\includegraphics[width=0.5\textwidth]{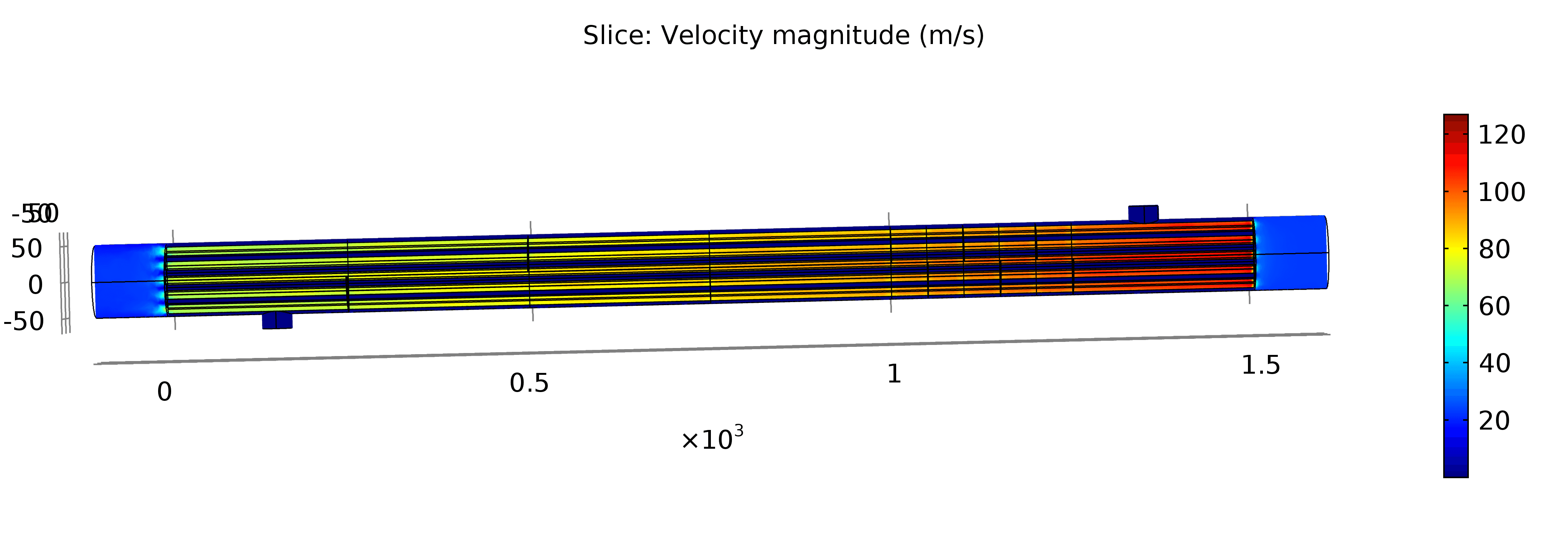}}
\subfloat[Temperature \label{subfig-2:dummy} ]{\includegraphics[width=0.5\textwidth]{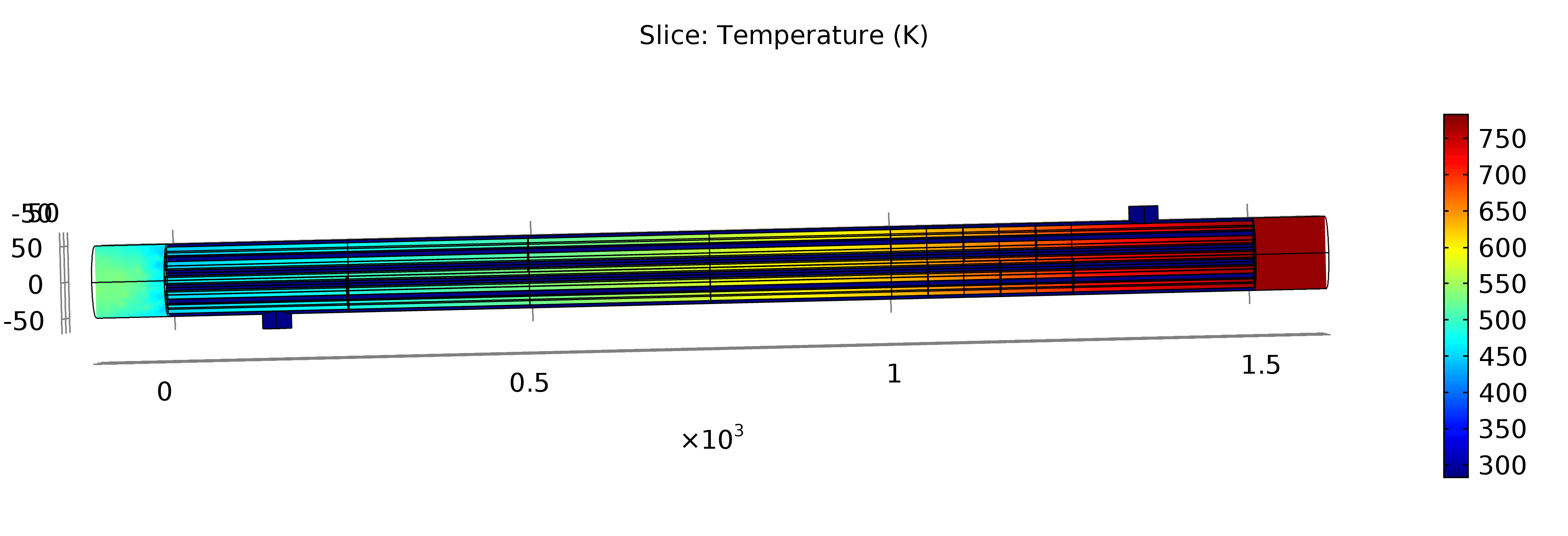}}
\hfill
\centering
\caption{Results for plain tubes with "pizza" baffles.}
\label{fig:dummy}

\end{figure}

 \begin{figure}[H]

\subfloat[Velocity \label{subfig-1:dummy}] {\includegraphics[width=0.95\textwidth]{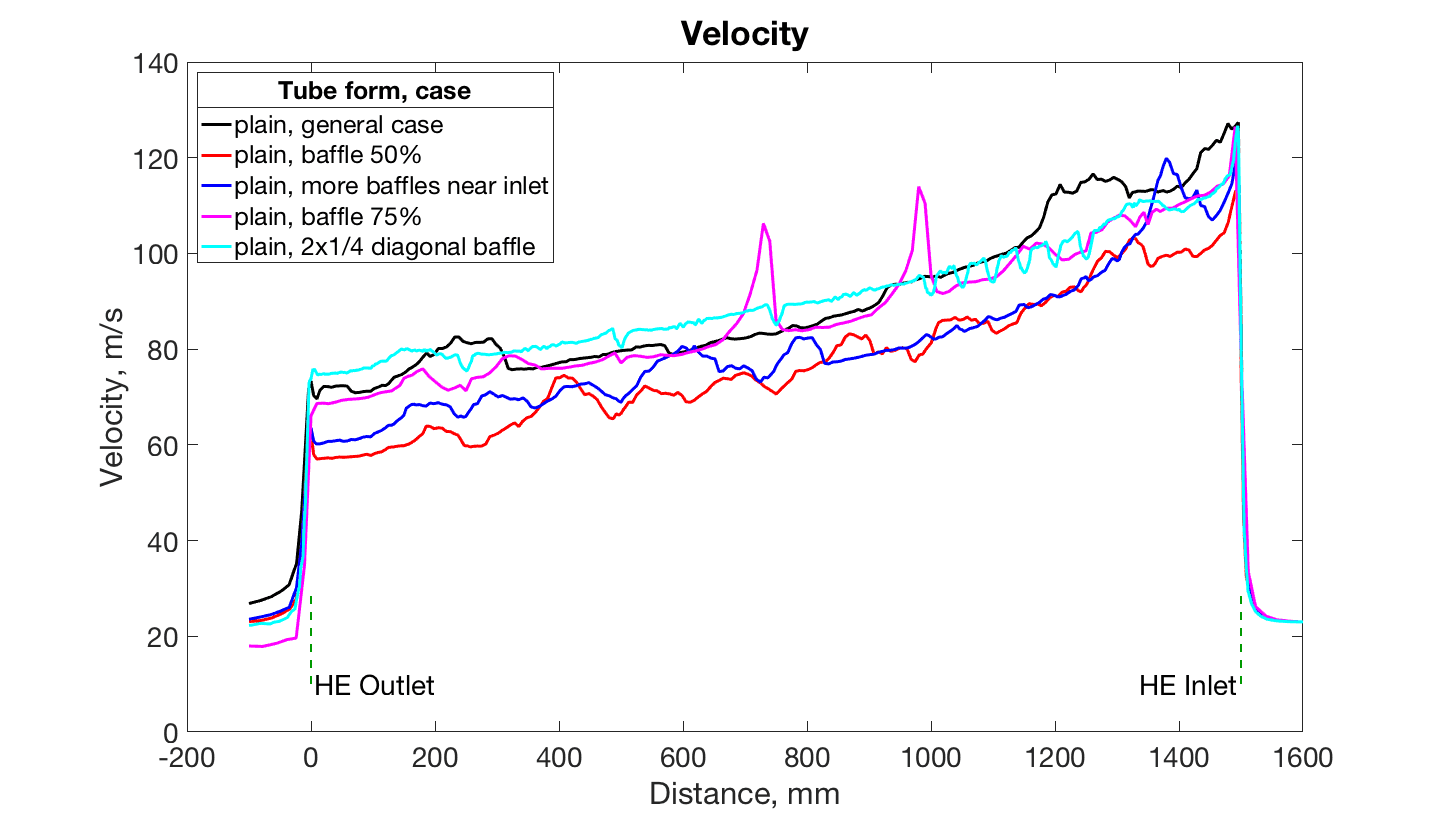}}
\hfill
\subfloat[Temperature \label{subfig-2:dummy} ]{\includegraphics[width=0.95\textwidth]{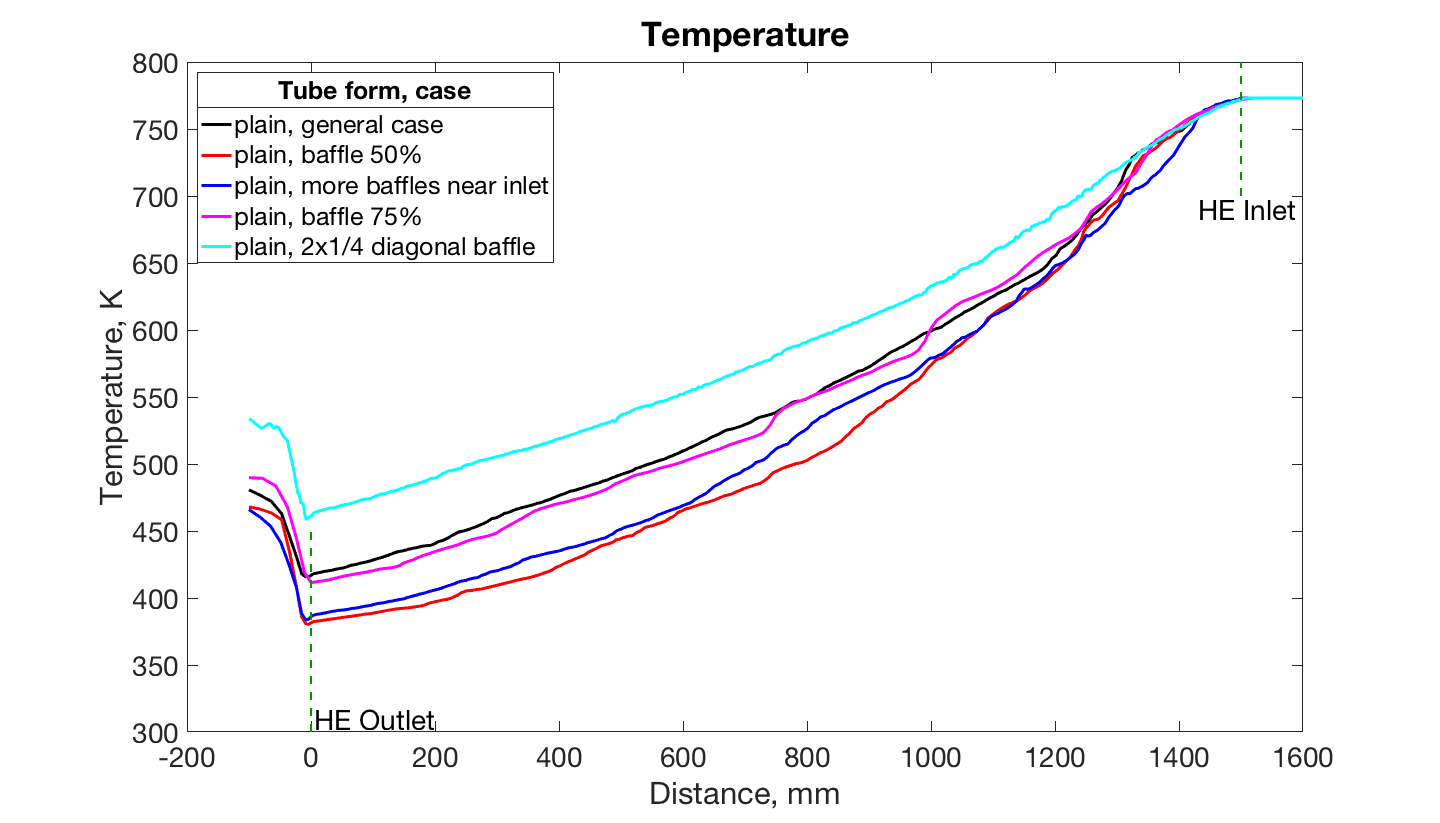}}
\hfill
\centering
\caption{Results for all five cases at section 3.4.}
\label{fig:dummy}

\end{figure}

Comparing red and black cases lower temperature along all tube length can be noticed with baffles. It is true as baffles increase turbulence inside a shell and heat transfer is higher. Outlet temperatures are: 467.98 K and 480.70~K for red and black models respectively. Velocity for general case is higher and outlet velocities are: 23.07 m/s and 26.84 m/s for red and black models respectively. \\
Model with more baffles near water inlet have the similar to red line graph. Outlet temperature is approximately the same as for half circle case -- 465.98~K. Small oscillations can be seen at the velocity figure at the range of 1000 -- 1350~mm. It is caused by mixing water flow near baffle at this location. This model has better efficiency than general case, but no much more than red case. \\
In case of plain tubes with circular segment as 3/4 of a circle areas with higher velocity near baffles can be noticed on the model graphs -- it is 'dead' zones that cause negative influence on the efficiency. Velocity peaks are due to these zones. Final temperature is higher -- 489.82~K. Also baffles are changed in upside-down way that cause different water flow direction.
Last case in this section with "pizza" arrangement of baffles shows the worst temperature distribution alone the whole heat exchanger length. Final temperature is 533.67~K. It happens because higher velocity of air in most location. Peaks show positions of baffles in the shell. Heat transfer is not intensive because of low water flow mixing. \\

\textbf{Summary} \medskip

Five models were developed with different baffles' geometry and location inside a shell. It was shown that cases with half circle baffles as non uniformly as regularly located have lower temperature and higher heat transfer efficiency. A model with a baffle as 3/4 circular segment of a circle has a bit higher temperature than general case because of dead zones and opposite disposition of baffles. "Pizza" form of baffles has the worst results both in velocity and temperature. \\

\subsection{Change of inlet \& outlet location and size}

In this section several models with changes in inlet and outlet of water are developed. Initially in general case both channels are located at 150 mm from edges of a heat exchange. First modification is to set water holes at a distance of 50 mm from the edges. Second way is to set inlet at 250 mm and outlet to 50 mm. At this case distance between holes remains the same as in general case. Next model has wider inlet and outlet for water -- 30 mm in radius instead of 20 mm in general case. The last model is with changes of air inlet and outlet. The length of channels is five diameters of a shell and equals 500 mm instead of 100 mm in general case. The results of a modelling are demonstrated on the Figures 15--17 and all graphs are presented on the Figure 18. 

\begin{figure}[H]

\subfloat[Velocity \label{subfig-1:dummy}] {\includegraphics[width=0.5\textwidth]{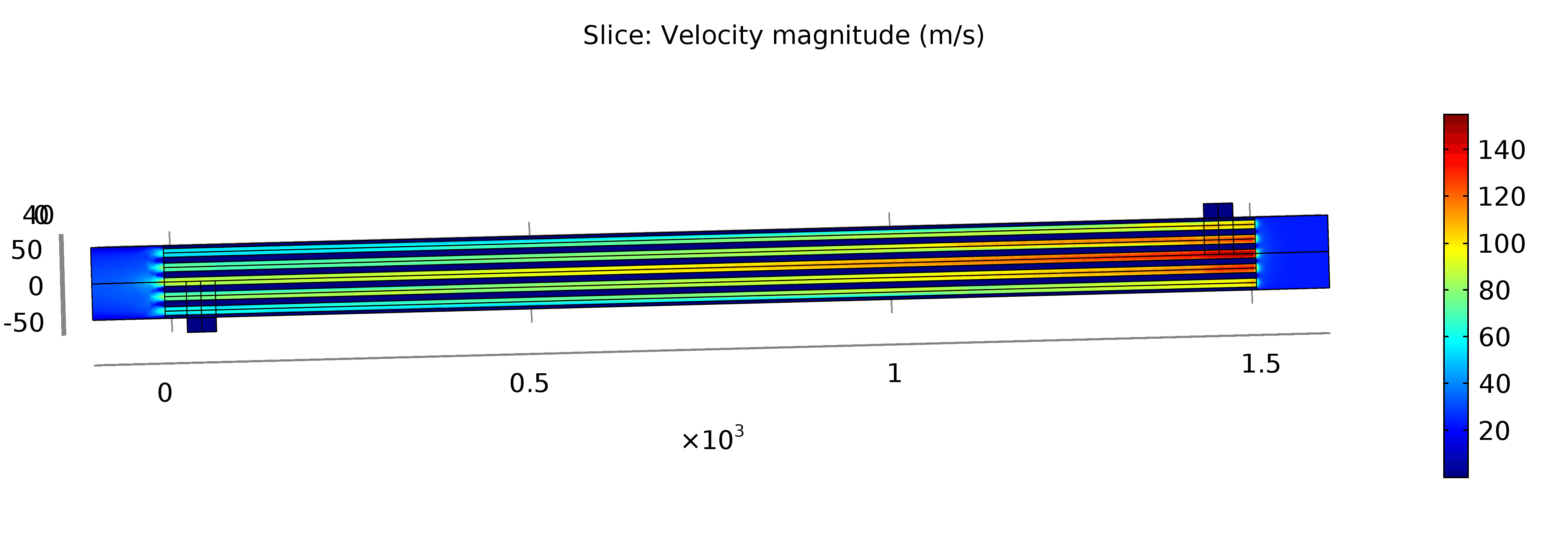}}
\subfloat[Temperature \label{subfig-2:dummy} ]{\includegraphics[width=0.5\textwidth]{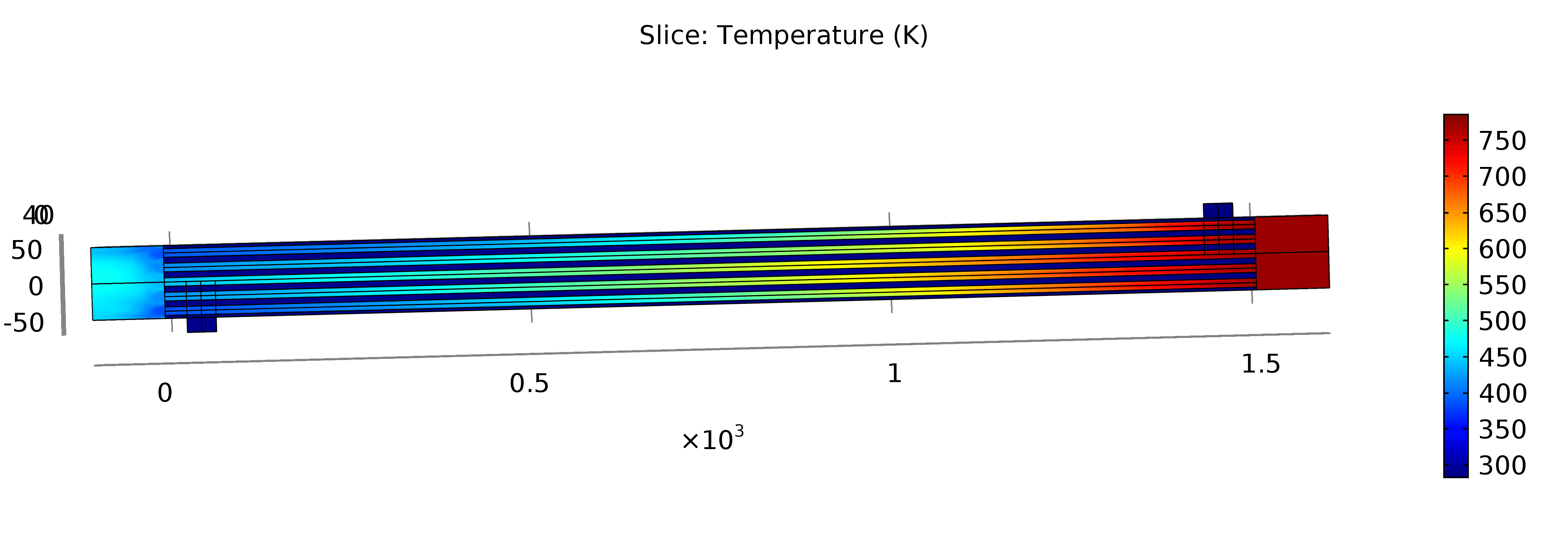}}
\hfill
\centering
\caption{Results for inlet 50 mm -- outlet 50 mm.}
\label{fig:dummy}

\end{figure}

\begin{figure}[H]

\subfloat[Velocity \label{subfig-1:dummy}] {\includegraphics[width=0.5\textwidth]{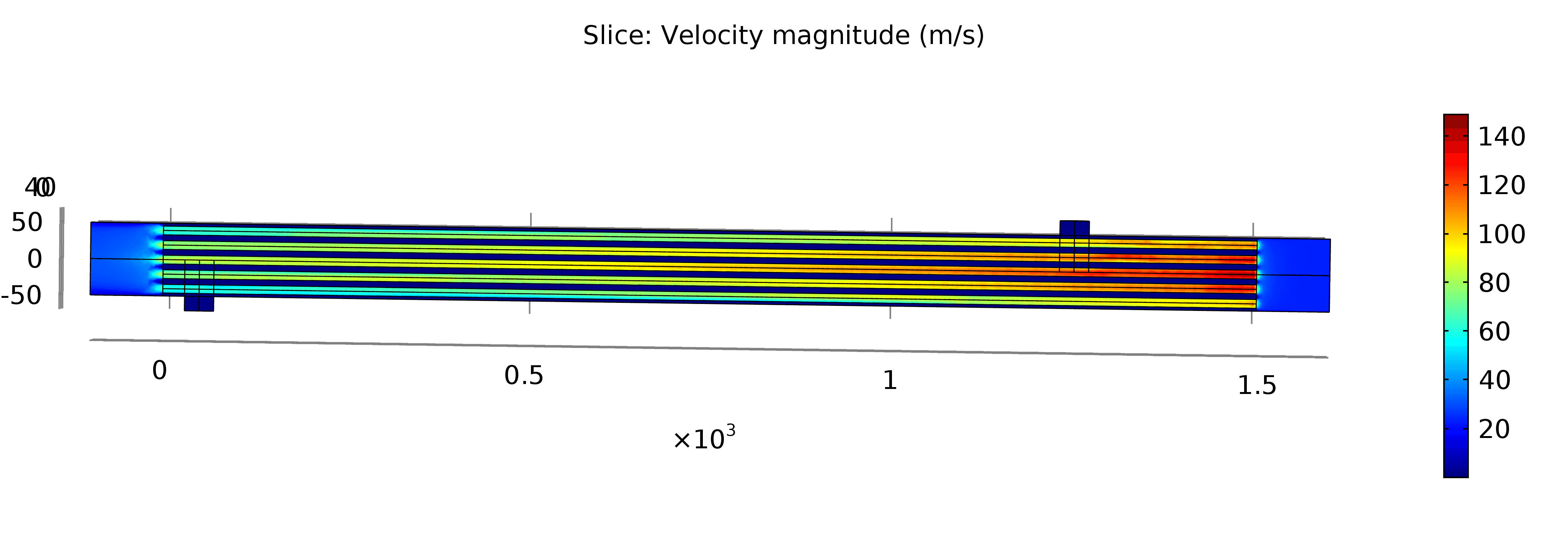}}
\subfloat[Temperature \label{subfig-2:dummy} ]{\includegraphics[width=0.5\textwidth]{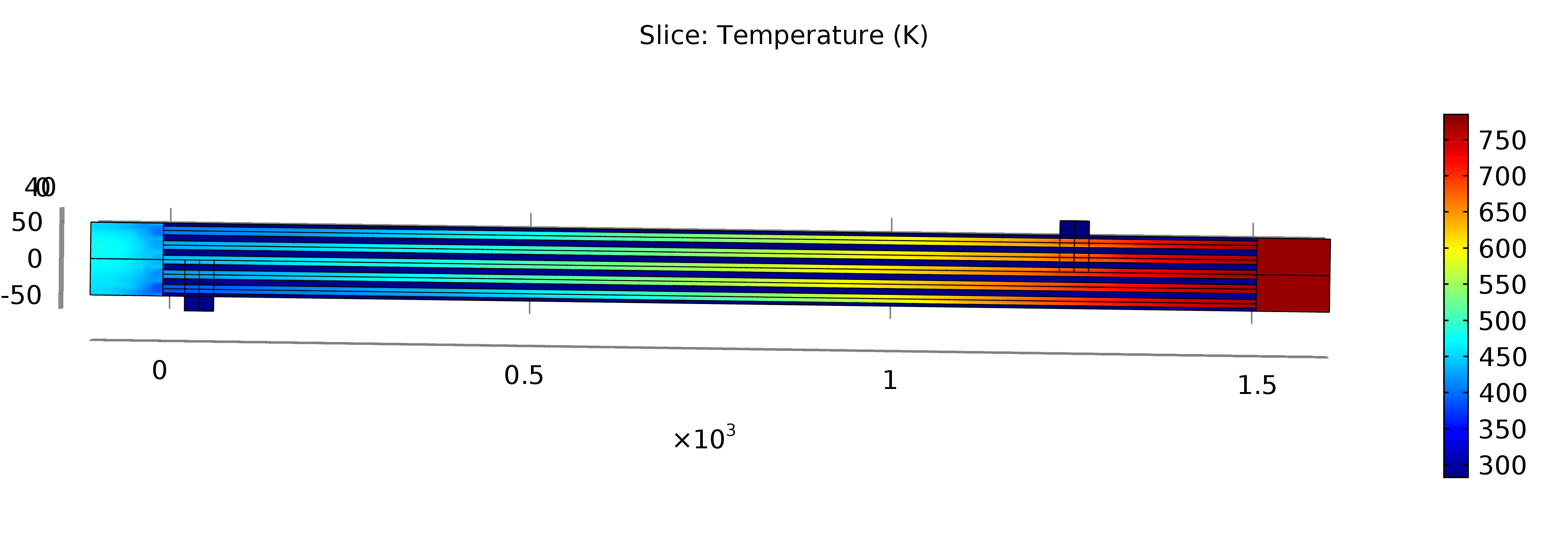}}
\hfill
\centering
\caption{Results for inlet 250 mm -- outlet 50 mm.}
\label{fig:dummy}

\end{figure}




 \begin{figure}[H]

\subfloat[Velocity \label{subfig-1:dummy}] {\includegraphics[width=0.75\textwidth]{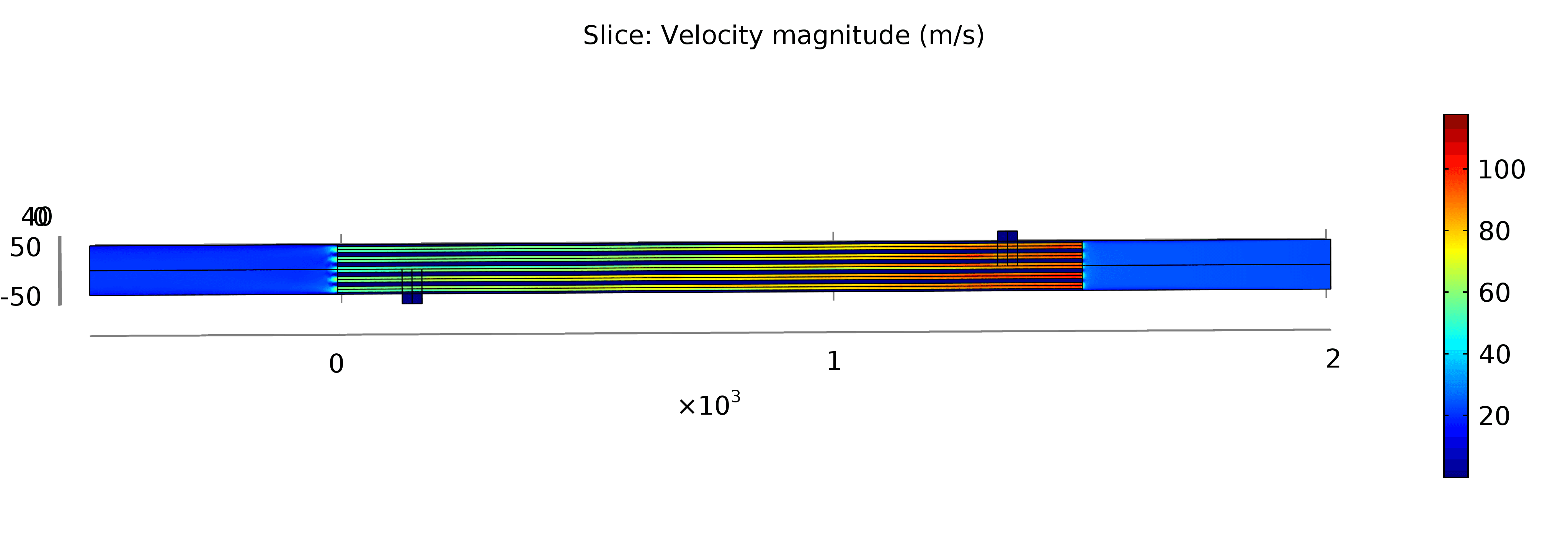}}
\hfill
\subfloat[Temperature \label{subfig-2:dummy} ]{\includegraphics[width=0.75\textwidth]{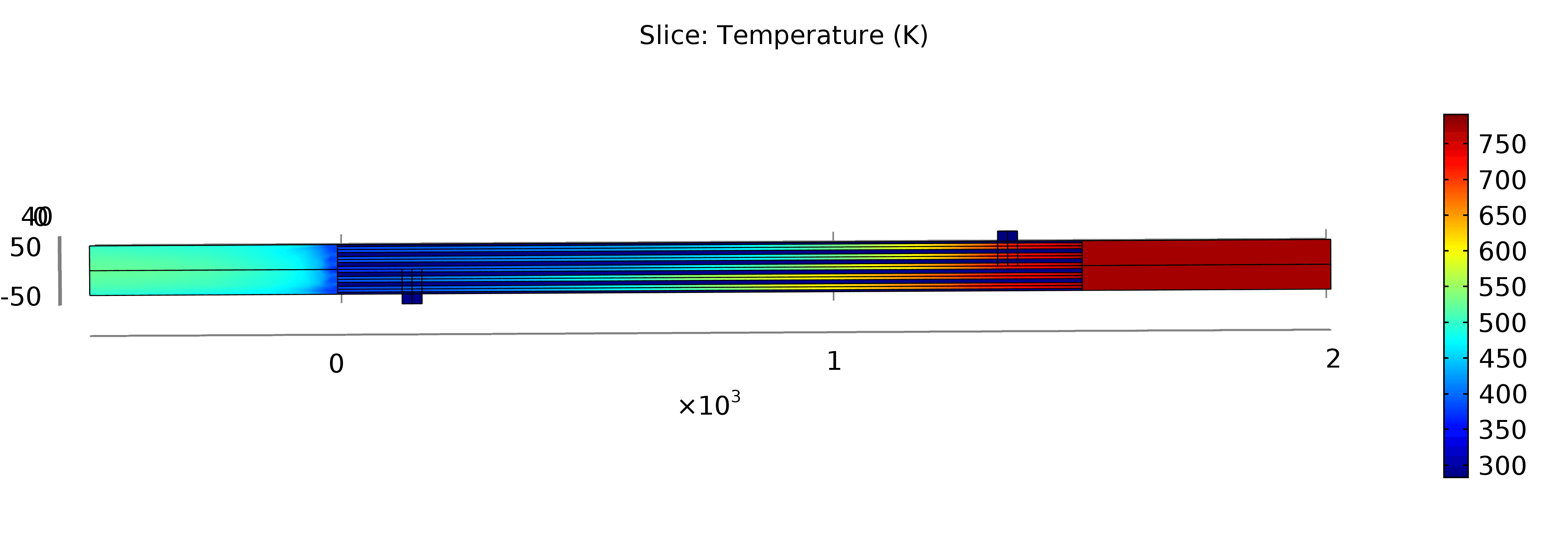}}
\hfill
\centering
\caption{Results for a long 500 mm air inlet \& outlet.}
\label{fig:dummy}

\end{figure}

 \begin{figure}

\subfloat[Velocity \label{subfig-1:dummy}] {\includegraphics[width=0.95\textwidth]{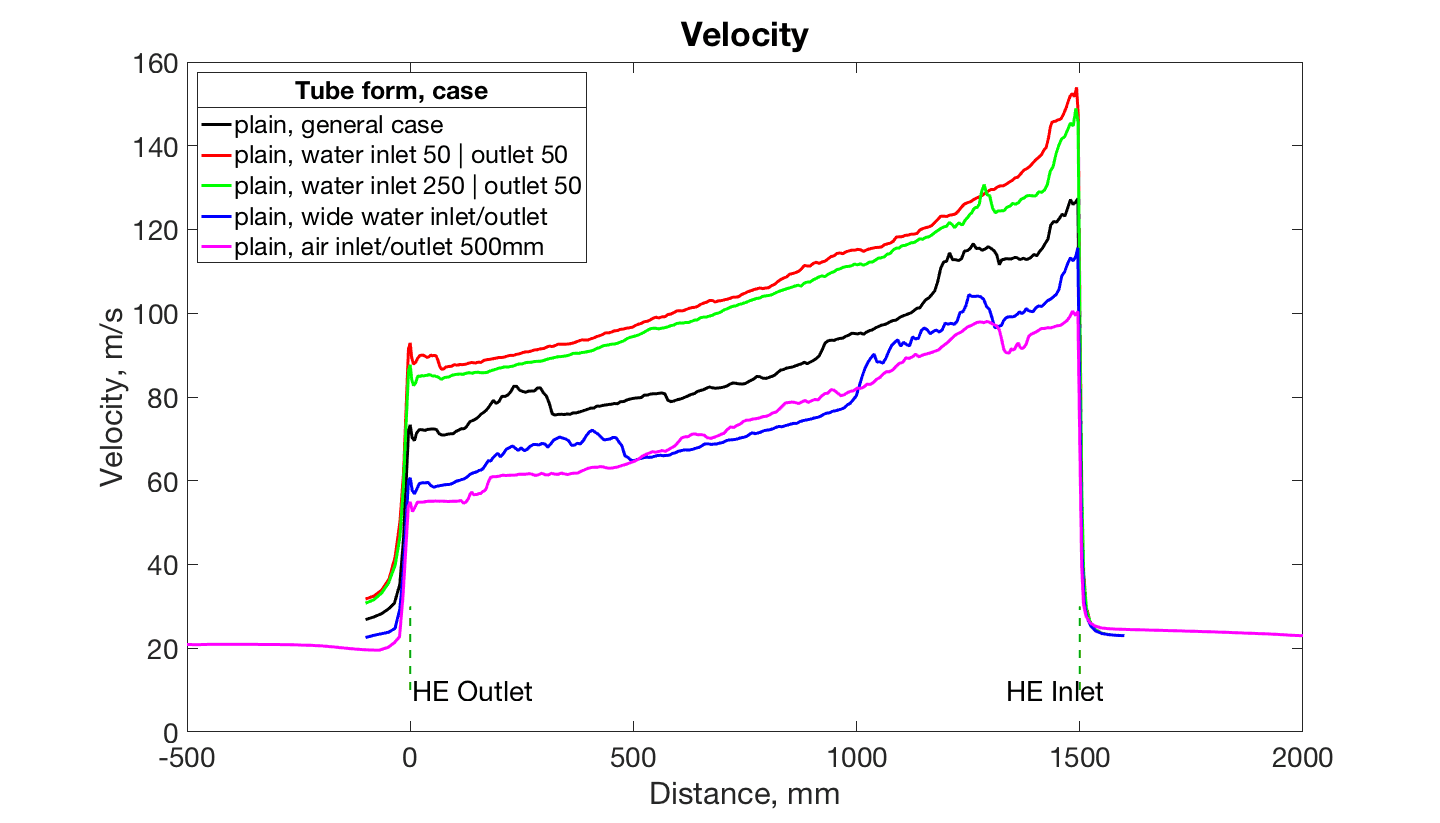}}
\hfill
\subfloat[Temperature \label{subfig-2:dummy} ]{\includegraphics[width=0.95\textwidth]{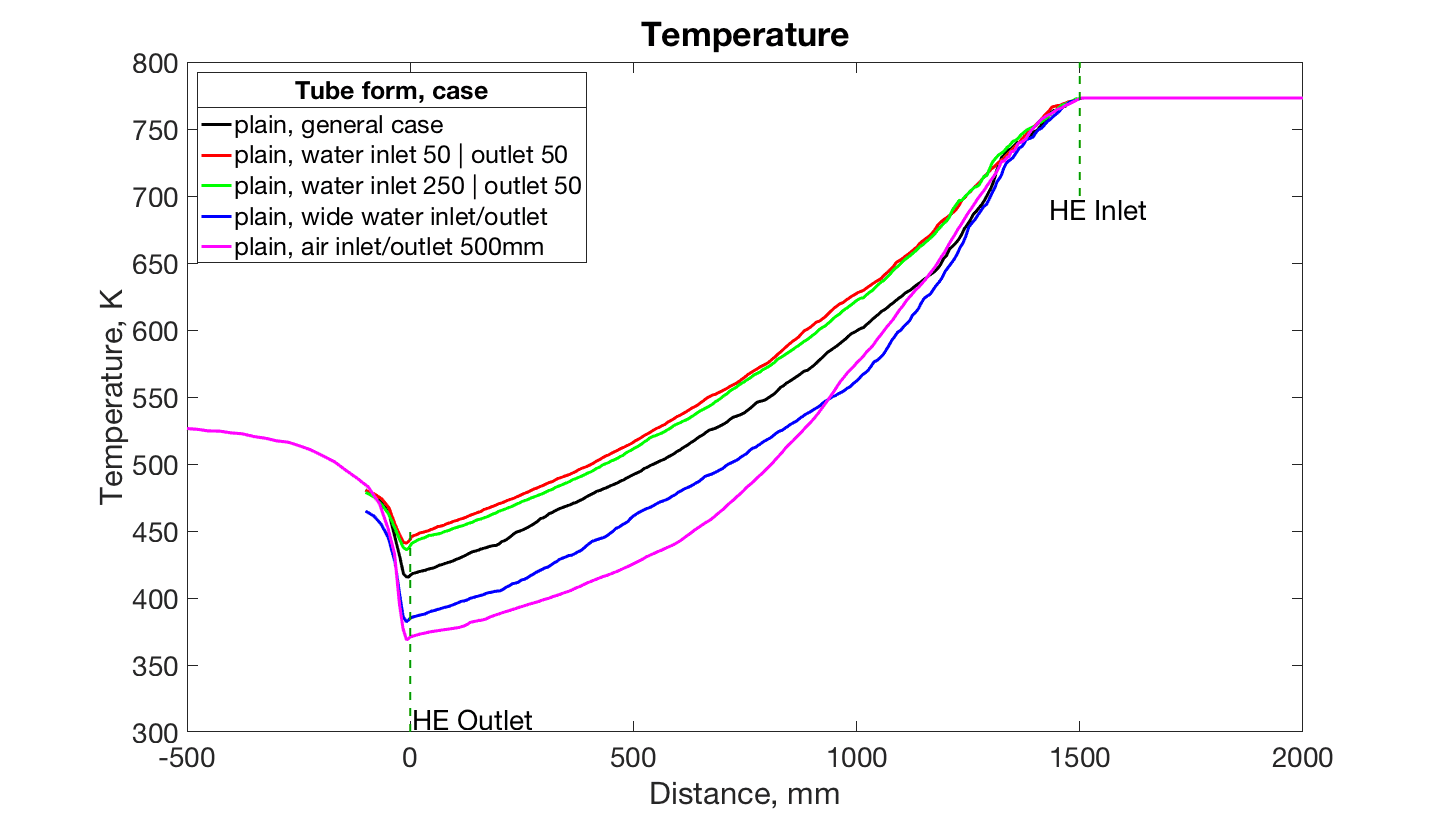}}
\hfill
\centering
\caption{Results for all five cases at section 3.5.}
\label{fig:dummy}

\end{figure}

Comparing red line, when water inlet and outlet are located in 50 mm from the edges, and black one higher temperature and velocity can be noticed in the first case. Outlet temperatures are: 480.19~K and 480.70~K for red and black models respectively. For general case the line is steeper at the end of plot, that means after mixing the flow temperature for red case should be lower farther. Velocity for general case is higher and outlet velocities are: 23.07 m/s and 26.84 m/s for red and black models respectively. It is seen that in the central tube for red model velocity is higher than for other tubes in a slice. This is due to the fact that water inlet is located closer to hot flange where air inlet is located. Central tube is washed not enough and has higher values of velocity and temperature. \par
Green plot looks quite similar to red one. Final temperature is 478.78~K that is also less than for general case. The explanation for this may be almost the same as for previous model -- central tube is washed not enough as it is located far away from the hot air inlet and has higher values of velocity and temperature. \par
Model with wide water inlet and outlet has lower value of temperature and velocity. Final temperature is 464.90~K and velocity -- 22.57 m/s. Larger diameter of channels at the same initial velocity and temperature leads to decrease of air parameters as more water flow washes tubes inside a shell. \par
The last model shows air flow that starts at a distance of 500 mm from a heat exchanger and ends at the same interval after a HE. It is created to look at flow distribution along the whole length as it is important where to measure parameters with sensors. At the velocity plot higher values can be noticed for general case than for magenta case. At the temperature graph huge drop inside a tube is observed. After HE outlet temperature changes from about 370~K to 530~K. All this changes occur due to air flow mixing before heat exchanger inlet and after outlet. It can be noticed that at a distance of 100 mm after HE outlet magenta line intersects line of a general case, so at this point temperatures are equal. \\

\textbf{Summary} \medskip

Five models were developed with different inlet and outlet locations and sizes. It was shown that changes in inlet hole closer or father lead to increase of temperature and velocity inside a central tube and has approximately the same temperature at a distance of 100 mm from an air outlet. Wide water channels at the same initial conditions decrease outlet air temperature and velocity. Long air inlet leads to changes of a flow and decrease temperature and velocity inside a tube. After a heat exchanger air flow is mixed a lot that causes rise in temperature.

\subsection{Additional models with different parameters}

Additional models were created taking different materials, tube's thickness. All models are with smooth tubes and without baffles inside a shell. In the first model air is replaced by steam. In the second case tube material is changed from stainless steel to AISI 4340 Alloy Steel. Tube's thickness was changed from 1mm to 2mm in the next model. And in the last case plain wires of 1 mm diameter are placed inside tubes.  General graphs for velocity and temperature are shown on the Figure 19.







It can be seen that blue, green lines and black asterisk line are similar on both temperature and velocity graphs. Final temperatures are: steam case -- 512.64~K, tube material change -- 480.67~K, tube thickness change -- 481.09~K,  case with wires inside tubes - 483.87~K, for general case -- 480.70~K.\\
Changing of tube material and tube thickness does not influence of the final temperature, as thermal conductivity coefficient and heat transfer has no big difference. Steam has lower thermal conductivity coefficient and higher temperature inside a tube as it should be. Putting inside a tube plain wires increase air velocity, as it does not add in mixing of a flow and  hydraulic diameter becomes less. The final temperature is a bit higher than in general case.

 \begin{figure}[H]


\subfloat[Velocity \label{subfig-1:dummy}] {\includegraphics[width=0.9\textwidth]{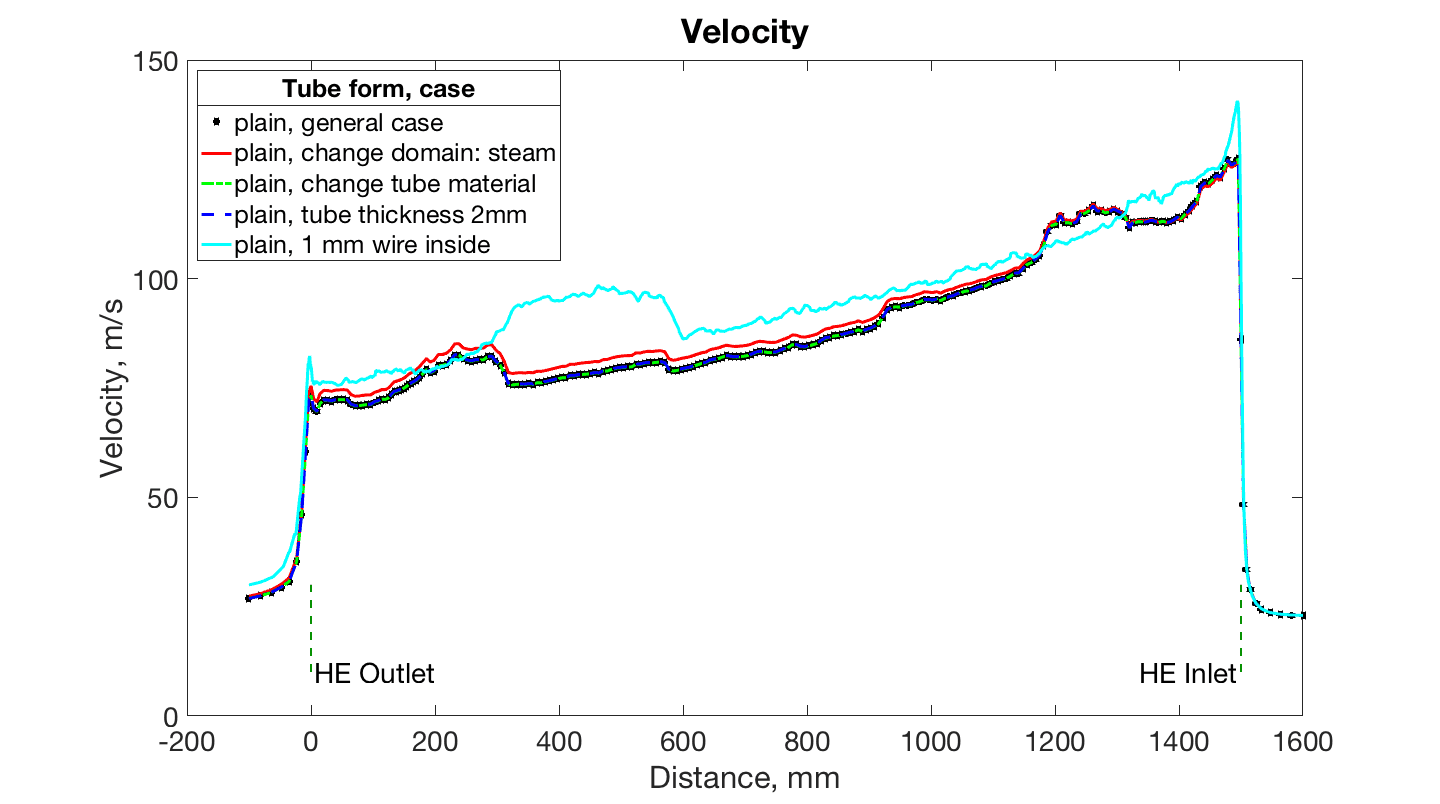}}
\hfill
\subfloat[Temperature \label{subfig-2:dummy} ]{\includegraphics[width=0.9\textwidth]{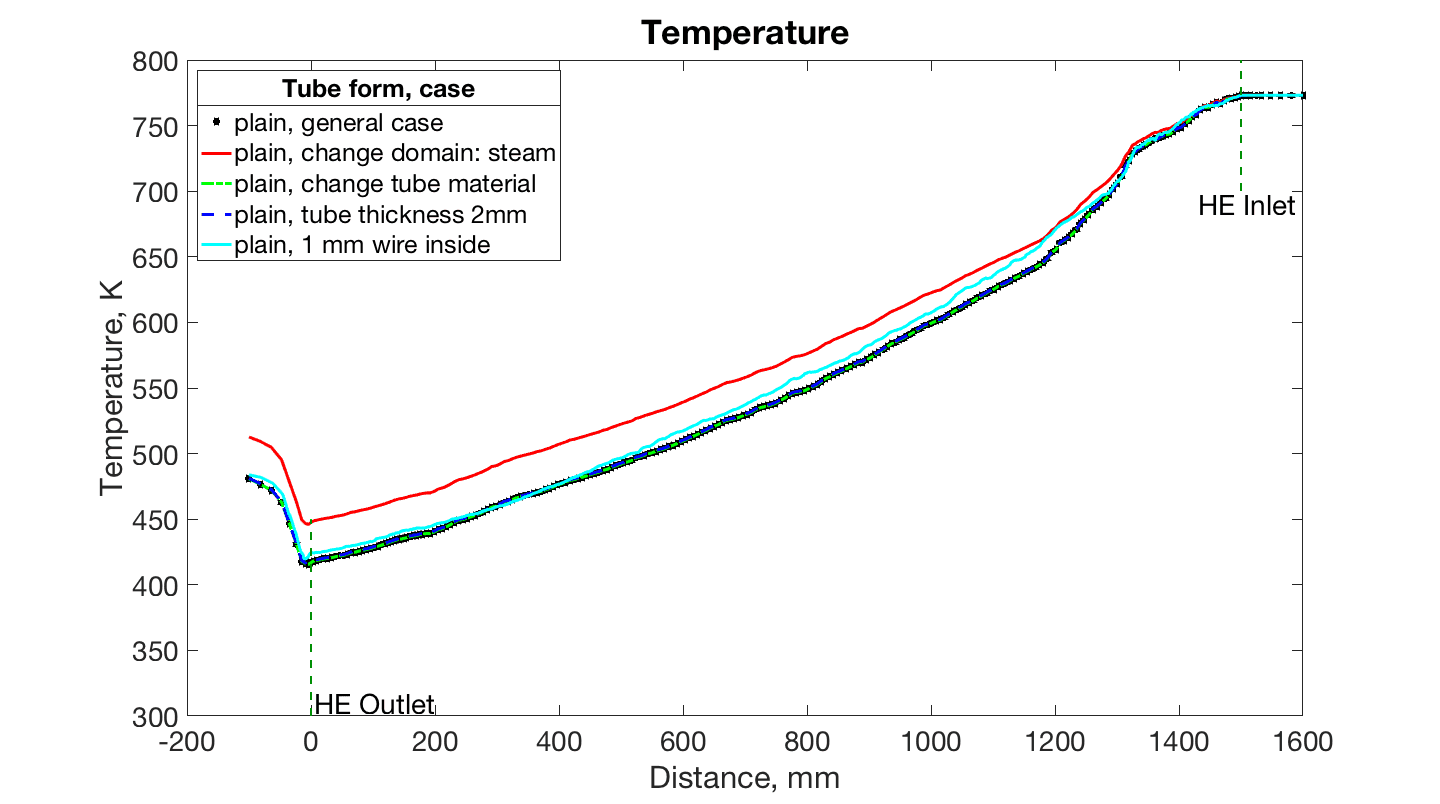}}
\hfill
\centering
\caption{Results for all five cases at section 3.6.}
\label{fig:dummy}

\end{figure}

\section{Experimental setup}

Several models were developed, computed and analyzed in the previous chapter. It can be seen that cases with twisted tubes have higher efficiency than with smooth tubes. To prove the conclusion of the simulation and have an opportunity to get experimental data from actual heat exchanger, apparatus with twisted tubes has been developed, experimental setup have been designed and the system is being created at Energy Systems Lab, Skoltech.  \\

\subsection{Design of a setup}

The setup is designed for testing different types of heat exchangers. One of them is for cars exhaust system. Considering this maximum temperature was taken as $500^{\circ}\mathrm{C}$ and air flow consumption of a heater at this temperature is 1500 L/min. All tubes are 100 mm in diameter. The scheme of the setup for gas-liquid case is demonstrated on the Figure 20.

\begin{figure}[H]
\includegraphics[width=0.7\textwidth]{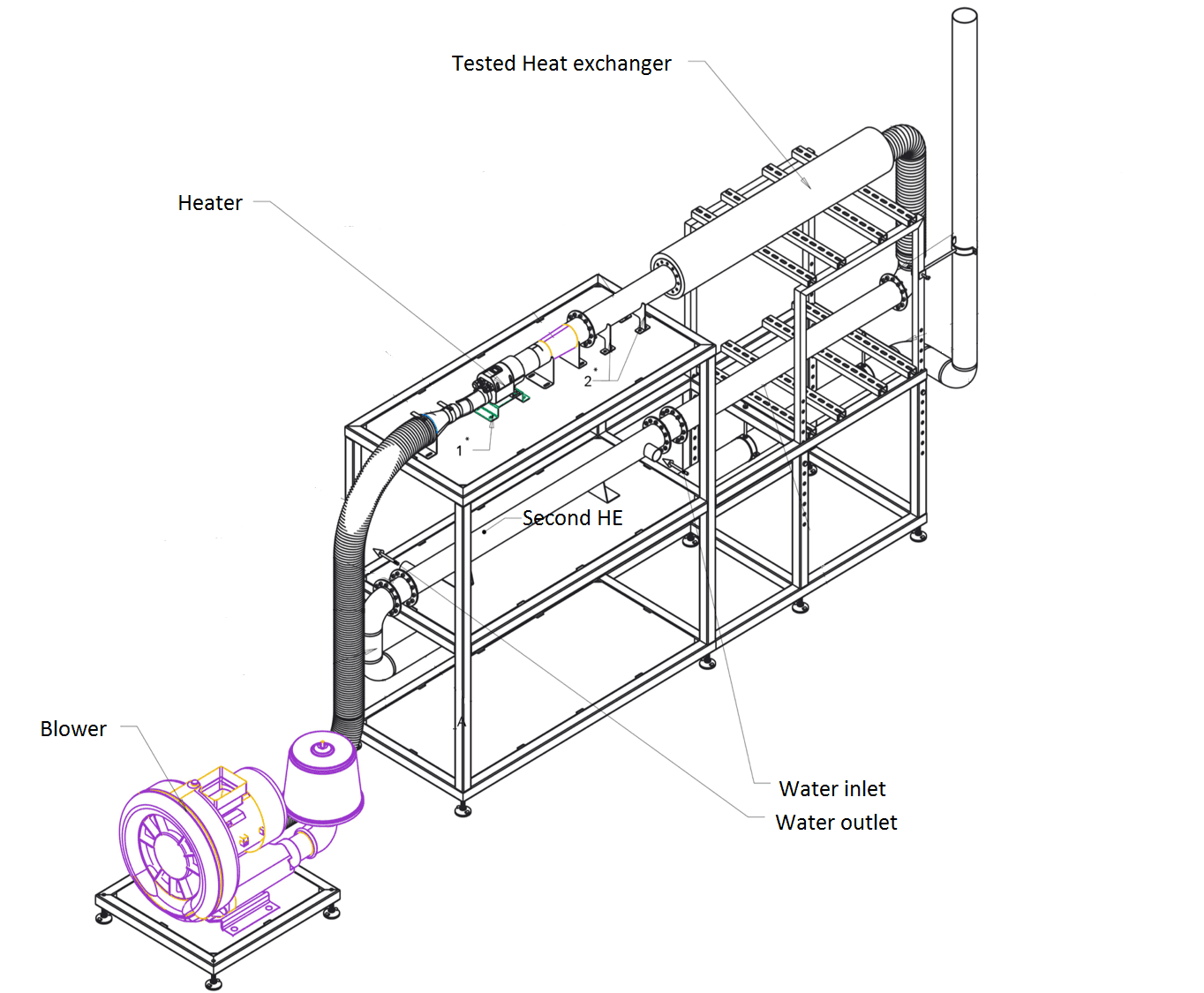}
\centering
\caption[Heat Exchanger]{Scheme of the setup for testing heat exchangers.
\label{fig:HE}}
\end{figure}

There are 3 levels in form of snake. Air blower FPZ SCL K10 is installed at the beginning (Figure 21 B), Maximum flow is 387 $m^3/h$. After corrugated pipe air goes to a heater Leister LHS 61S System 9kW (Figure 21 A). Then through a tube air goes through a first sensor zone to a tested heat exchanger. Heat exchanger with twisted tubes (Figure 21 C,D) inside was designed according the scheme and tube plate that is shown on Figure 22. After the exchanger second sensor zone is set, the same as the first one. Cooled air after the first line goes down to the second heat exchanger where it cools to the allowable temperature. It was developed this way to take into account such cases when tested heat exchanger can not cool air enough to throw out. Than second heat exchanger will cool down gas to the available temperatures. Third sensor zone is installed at the end to control exit temperature of the gas. At the end air goes on the third level to a gas hood. \\


\begin{figure}[H]

\subfloat[Heater \label{subfig-1:dummy}] {\includegraphics[width=0.45\textwidth]{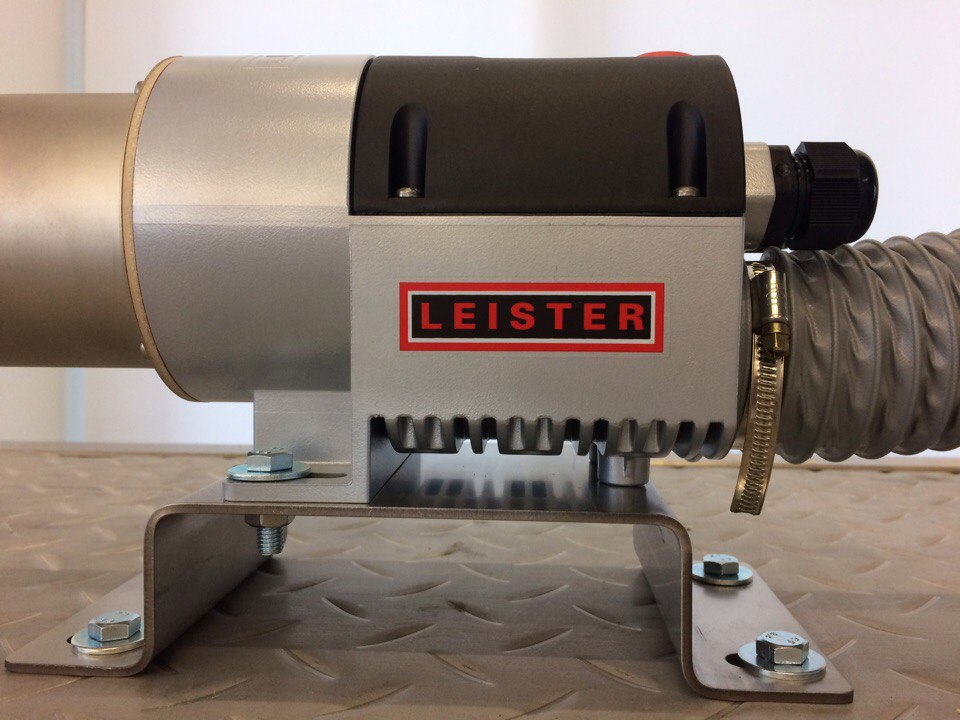}}
\hfill
\subfloat[Blower \label{subfig-2:dummy} ]{\includegraphics[ width=0.45\textwidth]{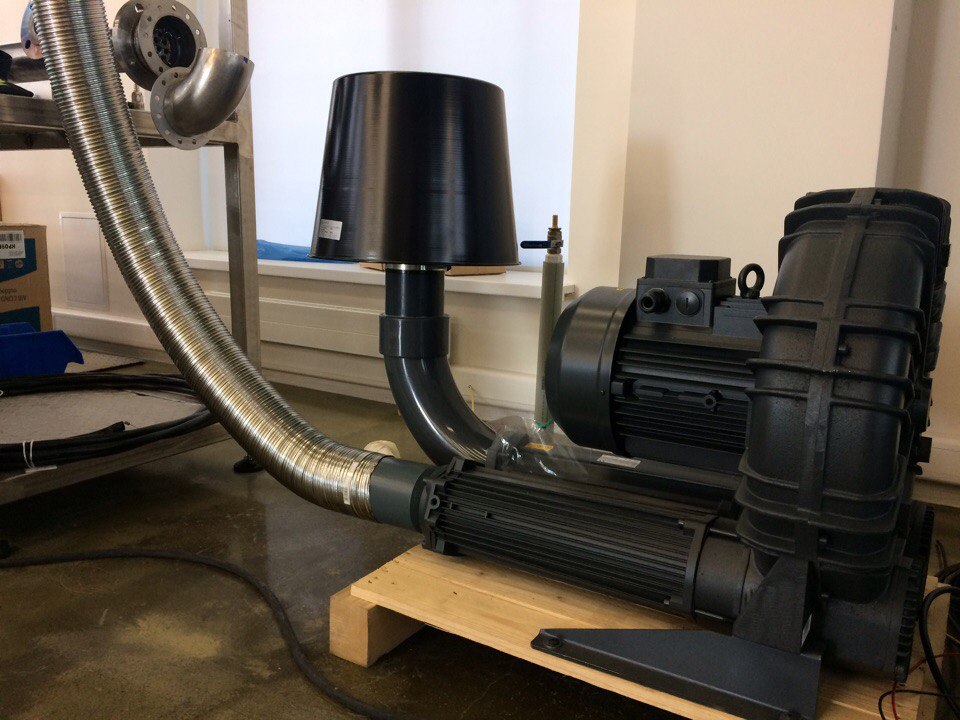}}
\hfill
\subfloat[HE flange. \label{subfig-2:dummy} ]{\includegraphics[height=6cm, width=0.45\textwidth]{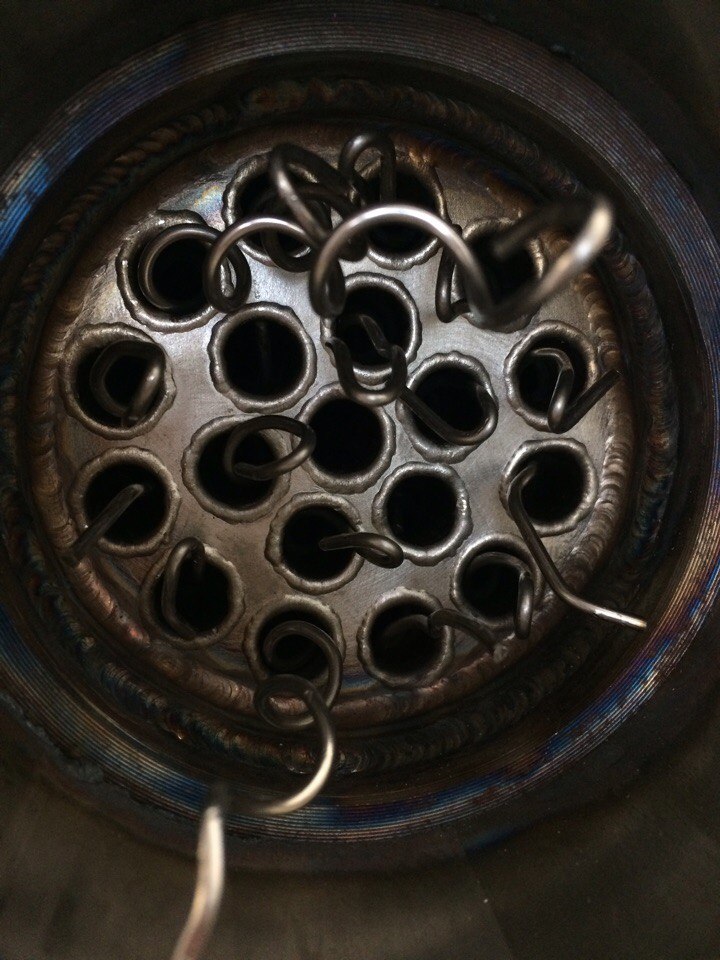}}
\hfill
\subfloat[Twisted tubes. \label{subfig-2:dummy} ]{\includegraphics[height=6cm, width=0.45\textwidth]{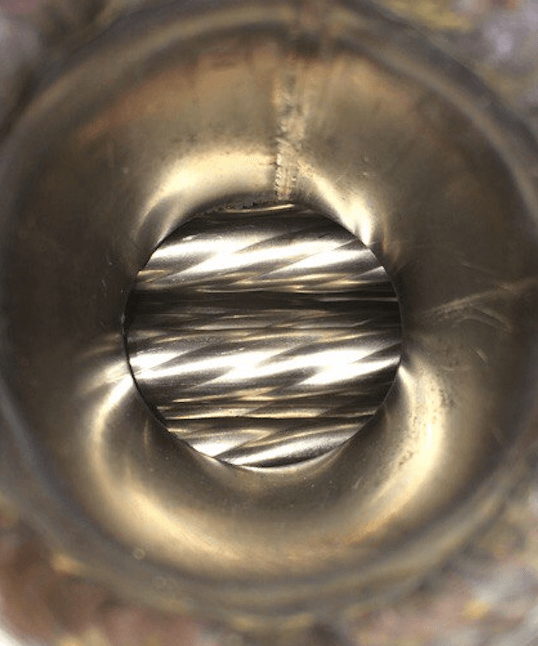}}
\hfill
\centering
\caption{Setup equipment.}
\label{fig:dummy}

\end{figure}

\begin{figure}[H]

\subfloat[Tube plate \label{subfig-1:dummy}] {\includegraphics[width=0.40\textwidth]{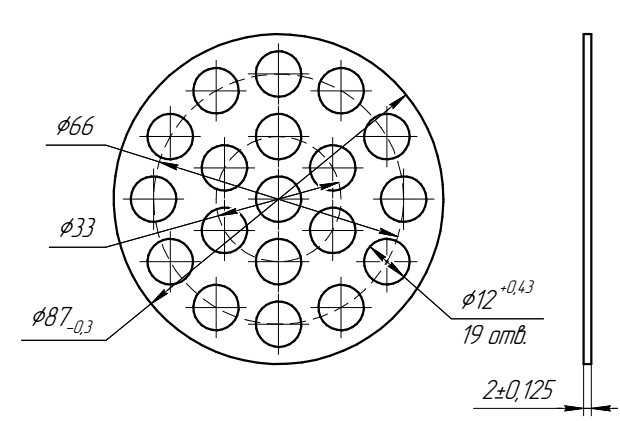}}
\hfill
\subfloat[Detailed scheme. \label{subfig-2:dummy} ]{\includegraphics[width=0.40\textwidth]{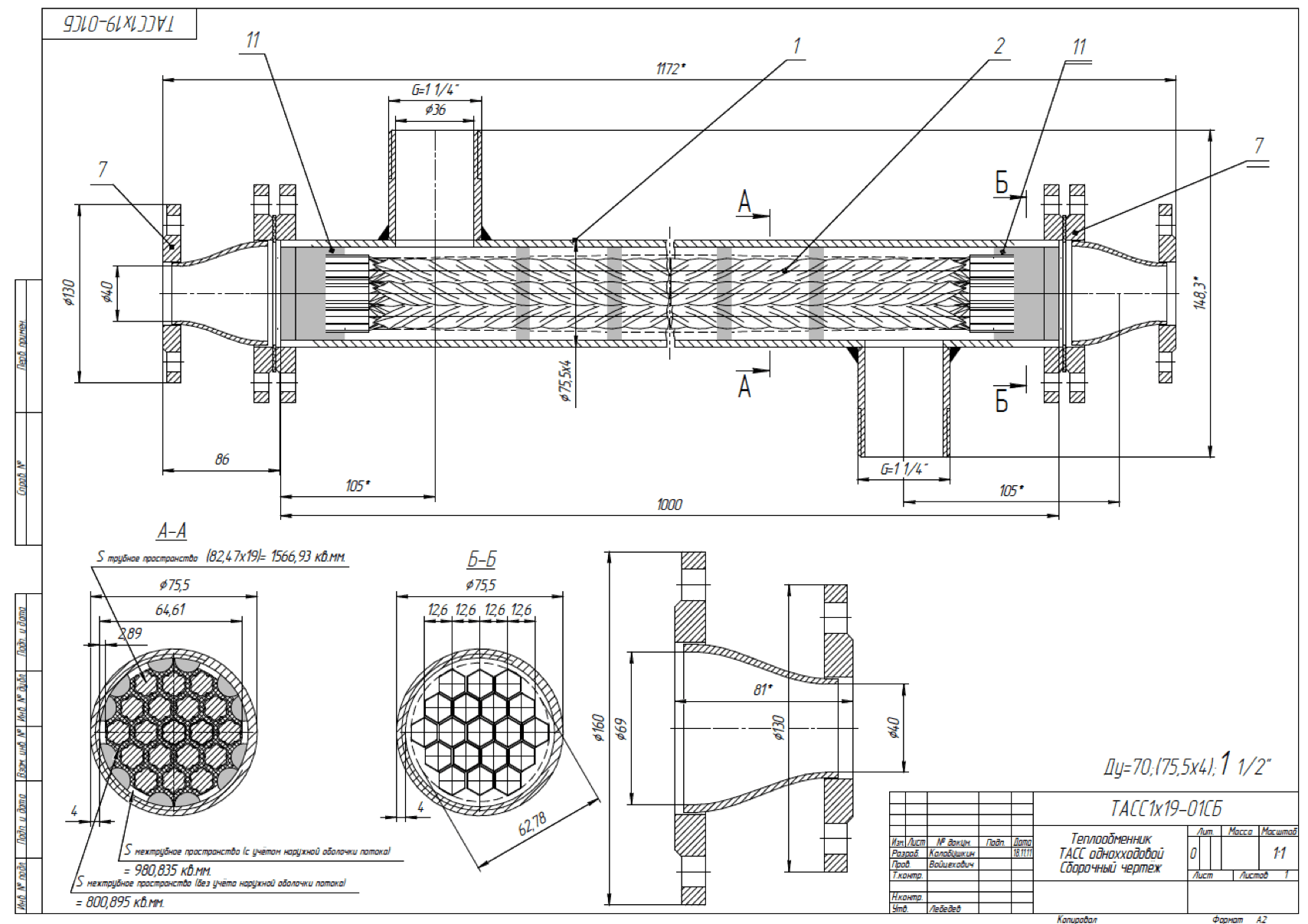}}
\hfill

\centering
\caption{Scheme of a heat exchanger.}
\label{fig:dummy}

\end{figure}

\subsection{Measurement system and techniques}

Sensor zone consists of Niiogas tube (Figure 23 B) with thermocouple, manometer model DMC-01M and a silicon tube for connection. Monometer measures temperature, pressure and flow rate and displays output data on the computer. For higher accuracy of measurements Coriolis flowmeter is used (Figure 23 A). \\

\begin{figure}[H]

\subfloat[Coriolis flowmeter \label{subfig-1:dummy}] {\includegraphics[width=0.4\textwidth]{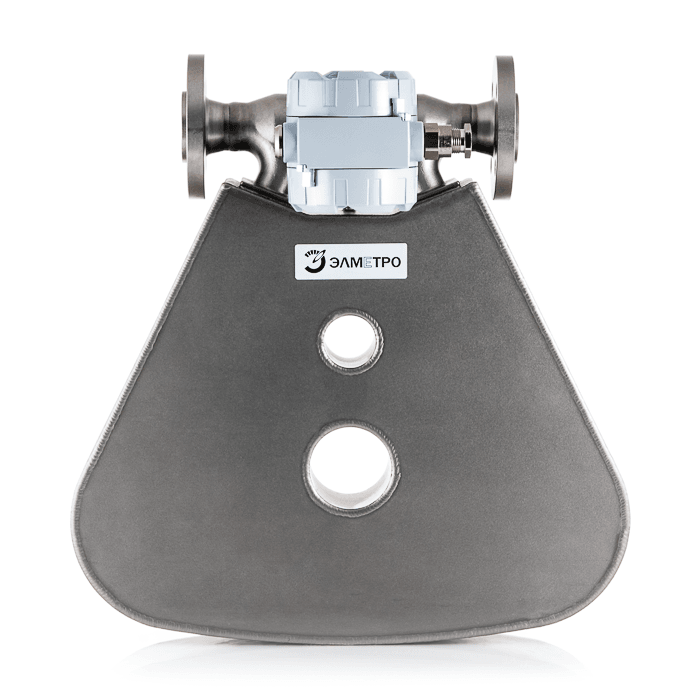}}
\hfill
\subfloat[Niiogas tube \label{subfig-2:dummy} ]{\includegraphics[width=0.40\textwidth]{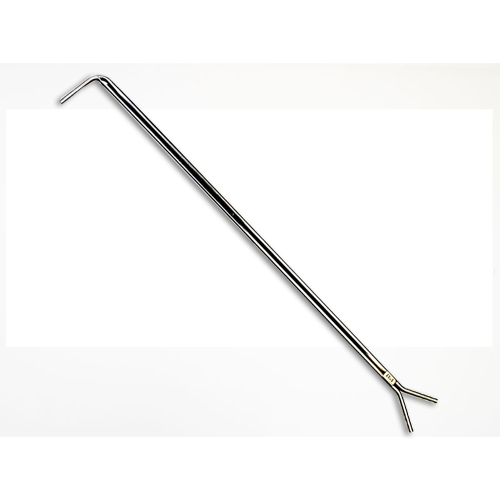}}
\hfill

\centering
\caption{Measurement equipment. }
\label{fig:dummy}

\end{figure}

Flowmeter El Metro Flomac: \\  
- measurement error: mass flow rate of gas: $\pm$ 0.35\%, medium temperature: $\pm$ 1.0 $^{\circ}\mathrm{C}$ \\
- measuring range: mass flow rate: from 1 kg/h to 540 t/h; Density: from 1 to 2000 kg / m3; Temperature of the medium to be measured: -60 to +350 $^{\circ}\mathrm{C}$. \\

Niiogas tube: \\ 
- measurement range of air (gas) flow rate, 2-60 m/s \\
- limiting range of operating temperatures,  -40 -- 600$^{\circ}\mathrm{C}$ \\
- relative error in determining the coefficient of the tube, not more than $\pm$5\% \\

Silicon tube has a temperature limit of 600$^{\circ}\mathrm{C}$, that is enough for the goal of the system.

Our experimental setup has several sensors for measurement improving. As was written above flowmeter has very high accuracy at the temperature below 350$^{\circ}\mathrm{C}$. But at higher temperatures in doesn't work properly. That is why tubes Niiogas are used to measure temperature, flow rate and pressure. Moreover, system with Niiogas tube cost mush less than Coriolis flowmeter and has higher temperature range. 

 There are several ways to measure thermal and hydraulic parameters. It can be measured directly with sensors. But it is not always possible. For example, there is no possibility to install some sensors. Then unknown parameters can be figure out of other data from equations that were presented in second chapter. \par
The whole system that is created on 05/2017 can be seen on the Figure 24. \\

\begin{figure}[H]
\includegraphics[width=0.6\textwidth]{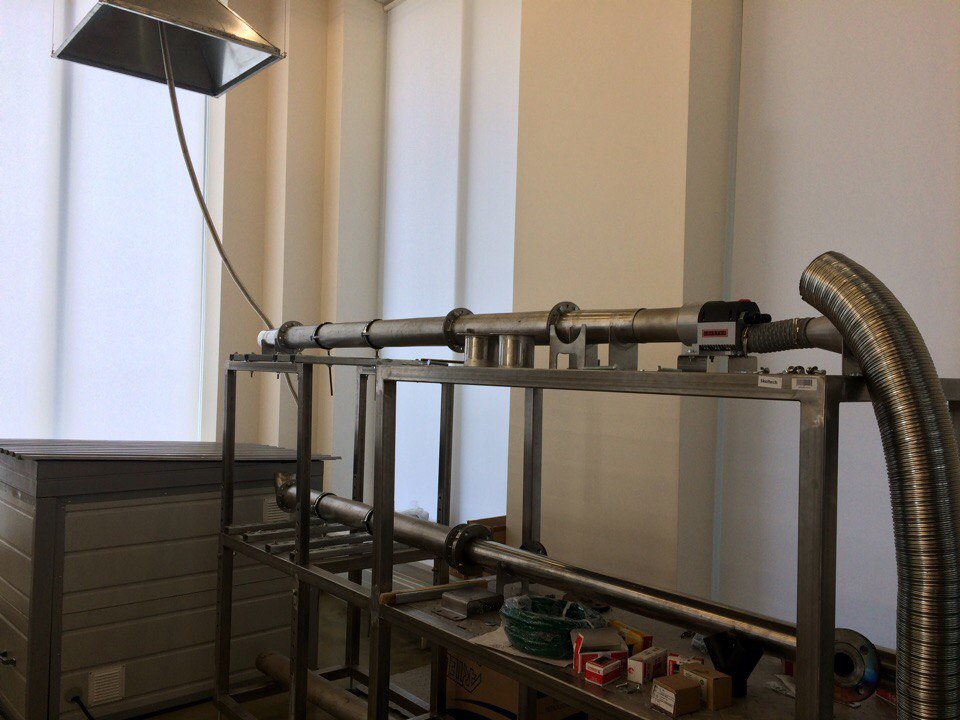}
\centering
\caption[Heat Exchanger]{Scheme of the setup for testing heat exchangers.
\label{fig:HE}}
\end{figure}


\section{Conclusion}
\addcontentsline{toc}{chapter}{Conclusion}

Theory on heat transfer and calculation of heat exchanger is presented in second chapter. According to it a program was created to compute shell and tube heat exchanger parameter from initial conditions. It is useful for experimental part to estimate what size, dimension and power HE will be if the inlet condition and material parameters are known. In the third chapter several CFD model are developed, computed and analyzed. Twisted tubes show significant increase in heat transfer efficiency. Setting baffles inside a shell leads to higher intensity of water mixing and lower temperature of a gas inside a tube. Construction with baffles should be without dead zones as it causes negative influence on the efficiency. Inlet and outlet location also contributes to the effectiveness of tube washing. Higher length of air inlet and outlet channels mixes flow and it is huge difference in temperature right after a heat exchanger and at a distance. This is important data for setting measurement equipment, as location has a value. It the fourth chapter experimental setup was presented and parameters of sensor systems.  \par
\break
\textbf{References} 









\begin{thebibliography}{1}
\addcontentsline{toc}{chapter}{Bibliography}


\bibitem{App1}  Heatric. Heat Exchanger Industry Applications. \url{http://www.heatric.com/heat\_exchanger}. 

\bibitem{App2} Applications of Heat Exchangers. \url{http://www.barriquand.com/en/heat-exchangers-industry}. 

\bibitem{SDH2}  Smart district heating and cooling, SETIS. \url{https://setis.ec.europa.eu/energy-research/publications/smart-district-heating-and-cooling}. 

\bibitem  {SDH1}  Large Scale Solar District Heating, PlanEnergi. \url{http://solar-district-heating.eu}. 

\bibitem{4712} Suhov V. V., Kazakov G. M. {\em Basics of designing and calculating heat exchangers}  2009. 


\bibitem{teplo2012} Gortishov U. F.  {\em Heat exchangers.}  2012.  






\bibitem{b0023} Pishchulin V. P. {\em Calculation of shell and tube heat exchanger}  2010. 

\bibitem{Isachenko} Isachenko V. P. {\em Heat Transfer, 3 edition} 1975. 





\bibitem{handbook} Kuppan Thulukkanam {\em Heat exchanger design handbook, second edition}  2013. 

\bibitem{koch}  Koch Heat Transfer Company. \url{http://www.kochheattransfer.com/products/twisted-tube-bundle-technology}. 

\bibitem{eddi} Eddify. \url{http://www.eddyfi.com/wp-content/uploads/2014/10/application-note-twisted-hx-tubes.pdf}. 

\bibitem{trin} Trinvalco. \url{http://www.trinvalco.com/Koch-Heat-Transfer-Company/twisted-tube-technology}. 

\bibitem{abi} ABI. \url{http://abi-us.com/faqs-twisted}.

\bibitem{1-s2} X. Yang  {\em Axial stiffness analysis of twisted tubes}, Procedia Engineering, 130, 298 -- 306, 2015. 

\bibitem{p1} Smith Eiamsa-arda, Pongjet Promvonge. {\em Performance assessment in a heat exchanger tube with alternate clockwise and counter-clockwise twisted-tape inserts.} International Journal of Heat and Mass Transfer, Volume 53, Issues 7--8, March 2010, Pages 1364--1372. 

\bibitem{p2} Smith Eiamsa-ard {\em Thermal characteristics in a heat exchanger tube fitted with dual twisted tape elements in tandem.} International Communications in Heat and Mass Transfer, Volume 37, Issue 1, January 2010, Pages 39--46. 

\bibitem{p3} Smith Eiamsa-ard {\em Experimental investigation of heat transfer and flow friction in a circular tube fitted with regularly spaced twisted tape elements.} International Communications in Heat and Mass Transfer, Volume 33, Issue 10, December 2006, Pages 1225--1233. 

\end{thebibliography}







\end{document}